\documentclass[fleqn,usenatbib]{mnras}

\usepackage{newtxtext,newtxmath}

\usepackage[T1]{fontenc}

\DeclareRobustCommand{\VAN}[3]{#2}
\let\VANthebibliography\thebibliography
\def\thebibliography{\DeclareRobustCommand{\VAN}[3]{##3}\VANthebibliography}

\usepackage{graphicx}	
\usepackage{amsmath}	
\usepackage[capitalise]{cleveref}
\usepackage{multirow}
\usepackage{threeparttable}
\usepackage{tabularx}
\usepackage{anyfontsize}
\usepackage{float}

\title[Excess ionospheric power in 21\nobreakdash-cm simulations]{Ionospheric contributions to the excess power in high-redshift 21\nobreakdash-cm power\nobreakdash-spectrum observations with LOFAR}

\author[S. A. Brackenhoff et al.]{
S.A. Brackenhoff,$^{1}$\thanks{E-mail: brackenhoff@astro.rug.nl (SAB)}
M. Mevius,$^{2}$
L.V.E. Koopmans,$^{1}$
A. Offringa,$^{2,1}$
E. Ceccotti,$^{1}$
J.K. Chege,$^{1}$
\newauthor
B.K. Gehlot,$^1$
S. Ghosh,$^1$
C. H\"ofer,$^1$
F.G. Mertens,$^{3,1}$
S. Munshi$^1$
and S. Zaroubi$^{4,1}$
\\
\\
$^{1}$Kapteyn Astronomical Institute, University of Groningen, PO Box 800, NL-9700 AV Groningen, The Netherlands\\
$^{2}$Netherlands Institute for Radio Astronomy (ASTRON), PO Box 2, NL-7990 AA Dwingeloo, The Netherlands\\
$^{3}$LERMA, Observatoire de Paris, PSL Research University, CNRS, Sorbonne Universit\'e, F-75014 Paris, France\\
$^4$ARCO (Astrophysics Research Center), Department of Natural Sciences, The Open University of Israel, 1 University Road, PO Box 808, \\Ra’anana 4353701, Israel
}

\date{Accepted 2024 July 29. Received 2024 July 29; in original form 2024 June 20}

\pubyear{2024}

\begin{document}
\label{firstpage}
\pagerange{\pageref{firstpage}--\pageref{lastpage}}
\maketitle

\begin{abstract}
The turbulent ionosphere causes phase shifts to incoming radio waves on a broad range of temporal and spatial scales. When an interferometer is not sufficiently calibrated for the direction-dependent ionospheric effects, the time-varying phase shifts can cause the signal to decorrelate. The ionosphere's influence over various spatiotemporal scales introduces a baseline-dependent effect on the interferometric array. We study the impact of baseline-dependent decorrelation on high-redshift observations with the Low Frequency Array (LOFAR). Datasets with a range of ionospheric corruptions are simulated using a thin-screen ionosphere model, and calibrated using the state-of-the-art LOFAR Epoch of Reionisation pipeline. For the first time ever, we show the ionospheric impact on various stages of the calibration process including an analysis of the transfer of gain errors from longer to shorter baselines using realistic end-to-end simulations. We find that direction-dependent calibration for source subtraction leaves excess power of up to two orders of magnitude above the thermal noise at the largest spectral scales in the cylindrically averaged auto-power spectrum under normal ionospheric conditions. However, we demonstrate that this excess power can be removed through Gaussian process regression, leaving no excess power above the ten per cent level for a $5~$km diffractive scale. We conclude that ionospheric errors, in the absence of interactions with other aggravating effects, do not constitute a dominant component in the excess power observed in LOFAR Epoch of Reionisation observations of the North Celestial Pole. Future work should therefore focus on less spectrally smooth effects, such as beam modelling errors.
\end{abstract}



\section{Introduction} \label{sec:intro} 
Detecting the spatial variations in the redshifted 21\nobreakdash-cm line from neutral hydrogen during the Epoch of Reionisation (EoR) and Cosmic Dawn is an important objective in observational cosmology, because it offers a unique glimpse into the evolution of the early universe, its radiating sources, and how they interact with their environment (e.g. \citealt{furlanetto_cosmology_2006, morales_reionization_2010, pritchard_21-cm_2012, zaroubi_epoch_2013}). Inhomogeneity in the density of neutral hydrogen in the early universe causes the 21\nobreakdash-cm signal to vary spatially. The aim of interferometric  experiments is to capture these spatial variations in a power spectrum. 
However, achieving a detection of this high-redshift 21\nobreakdash-cm field remains technically challenging for many reasons (see \citealt{liu_data_2020}, for example). A number of low-frequency arrays have been built to detect this 21-cm signal over the past two decades, or are currently under construction. Arrays that have set upper limits on the 21\nobreakdash-cm power spectrum are
 the GMRT (Giant Metrewave Radio Telescope, \citealt{swarup_giant_1990,paciga_simulation-calibrated_2013})\footnote{\url{http://gmrt.ncra.tifr.res.in}}, 
 the MWA (Murchison Widefield Array, \citealt{bowman_science_2013,trott_deep_2020,barry_improving_2019, li_first_2019})\footnote{\url{http://www.mwatelescope.org}}, 
 PAPER (Donald C. Backer Precision Array to Probe EoR, \citealt{parsons_precision_2010,kolopanis_simplified_2019})\footnote{\url{http://eor.berkeley.edu}},
 HERA (Hydrogen Epoch of Reionization Array, \citealt{deboer_hydrogen_2017,abdurashidova_first_2022})\footnote{\url{http://reionization.org}}, 
 LOFAR (Low-Frequency Array, \citealt{haarlem_lofar_2013,patil_upper_2017,mertens_improved_2020})\footnote{\url{http://www.lofar.org}}, 
 OVRO-LWA (Owen's Valley Radio Observatory -- Long Wavelength Array, \citealt{hallinan_notitle_2015,eastwood_21_2019,garsden_21-cm_2021}),
 and NenuFAR (New Extension in Nan\c{c}ay Upgrading LOFAR, \citealt{zarka_lssnenufar_2012,munshi_first_2024})\footnote{\url{https://nenufar.obs-nancay.fr}}. 
 Furthermore, the construction of SKA (Square Kilometre Array, \citealt{dewdney_square_2009,koopmans_cosmic_2015})\footnote{\url{http://www.skatelescope.org}}, an instrument that will have a much higher sensitivity than its predecessors, has begun.\par
Galactic and extra-galactic foreground sources that are several orders of magnitude brighter obscure the 21\nobreakdash-cm signal \citep{shaver_can_1999}. However, the foregrounds are more spectrally smooth than the EoR signal, such that the two can be separated during data analysis. This requires high-precision calibration of the instrument. Issues can arise, for example, due to sky model errors and incompleteness \citep{ewall-wice_impact_2017}, or because the gain errors vary too fast as a function of time, direction, and baseline to be solved for. In addition, effects such as correlated radio-frequency interference \citep{offringa_impact_2019} and instrumental leakage between Stokes components or cable reflections \citep{jelic_realistic_2010, gasperin_effect_2018, offringa_precision_2019} can leave residuals in the data after calibration. Finally, there are ionospheric errors that introduce chromatic errors \citep{koopmans_ionospheric_2010,vedantham_scintillation_2016}. Combined, these effects lead to an `excess variance' that biases the recovered 21\nobreakdash-cm power spectrum \citep{mertens_improved_2020}. The origins of this excess variance have remained hard to identify (see e.g. \citealt{gan_assessing_2023}). The goal of this work is to assess the part that the ionosphere plays in this excess variance.\par
Recent work using LOFAR and NenuFAR, suggests that the excess variance may partially be attributed to residual foregrounds from bright off-axis sources (Ceccotti et al., in preparation) and to unflagged low-level Radio-Frequency Interference (RFI) emitted near the interferometer \citep{munshi_first_2024}. The former poses a challenge because the primary beam of the instrument is highly spatially, spectrally and to a lesser degree temporally varying far from the target direction (see e.g. \citealt{yatawatta_initial_2013, cook_calibration_2021}). Bright sources moving through side lobes in this variable beam cause an excess variance in the data, because of the inability to model the beam with sufficient accuracy \citep{asad_polarization_2018}. The difficulty of modelling this excess flux is further exacerbated by stochastic corruptions of the flux of these bright sources in the ionosphere (e.g. \citealt{koopmans_ionospheric_2010,2015MNRAS.453..925V}). It is of great importance to identify the impact of each possible excess variance source on the power spectrum, such that they may be mitigated. \par
The ionosphere is a known source of difficulty in creating radio-astronomical images, and many methods for correcting for the ionosphere have been investigated (see, for example, \citealt{intema_ionospheric_2009,weeren_lofar_2016,rioja_leap_2018,gasperin_reaching_2020,albert_probabilistic_2020,tasse_lofar_2021} and \citealt{chege_optimising_2022}). The ionosphere can affect the data in various ways, such as through its own thermal emission, refraction and diffraction, and in its lower layers also through absorption. While thermal emission and absorption is of great concern in global 21\nobreakdash-cm experiments \citep{sokolowski_impact_2015,datta_effects_2016, shen_bayesian_2022}, an interferometer is more sensitive to refraction and diffraction (depending on the lengths of its baselines). In the context of the foreground-avoidance approach followed by HERA, work has been done on estimating the ionospheric attenuation on polarised foregrounds \citep{martinot_ionospheric_2018}. The MWA is an instrument that is relatively similar to LOFAR. However, during calibration, the MWA uses only its shorter baselines, and assumes that the effect of the ionosphere can be corrected with source position-dependent refractive shifts \citep{lonsdale_configuration_2005, chege_simulations_2021} if the most active ionospheric time-bins are discarded \citep{trott_assessment_2018,waszewski_measurement_2022}. Work on MWA EoR measurements by \citet{jordan_characterization_2017} and \citet{chege_optimising_2022} has shown that the refractive effects of the ionosphere are not a dominant contributor to their excess variance and that updating the model based on ionospheric refraction does not improve results. Because the LOFAR calibration strategy uses longer baselines for calibration, the variation of ionosphere across the array cannot be ignored. This can cause sources to scintillate as well as refract, which has not been simulated in the context of MWA. \par
Analytic work on the expected variance of measured visibilities has been done by \citet{2015MNRAS.453..925V}. In a subsequent paper, \citet{vedantham_scintillation_2016} have computed the impact of scintillation noise on power spectra with an idealised source flux density distribution. However, the LOFAR EoR calibration strategy employs a method in which a longer subset of baselines is used for calibration and a shorter subset for power-spectrum analysis. This separation between baseline lengths is not considered in the work by \citet{vedantham_scintillation_2016}. \par
In this work, we present simulations of ionospheric effects on LOFAR EoR measurements taken with the High-Band Array (HBA), in which we can control the behaviour of the ionosphere and separate it from other in-situ effects. This provides a novel result, because the longer baselines of LOFAR cause us to be in a different ionospheric regime than earlier MWA-based work. Furthermore, we analyse whether ionospheric calibration errors from longer baselines are transferred to shorter baselines. We present, for the first time, a full simulation of ionospheric impact on the LOFAR EoR pipeline and assess the impact on the final power spectrum.\par
This work is structured as follows. First, we discuss the effect of the ionosphere on a minimally redundant array such as LOFAR in \cref{sec:iono}. In \cref{sec:methods}, we describe our end-to-end simulation and calibration pipelines. In these, we first generate data with ionospheric corruptions and then process this data in a pipeline that replicates the LOFAR EoR data processing pipeline. In \cref{sec:results}, we show the resulting power spectra, study the impact of the ionosphere on the gain solutions, and discuss how we can extrapolate the results to longer observations. In \cref{sec:disc}, we discuss the main conclusions of this work.
\section{The Ionosphere} \label{sec:iono}
The ionosphere is a partially ionised layer of turbulent plasma in the upper atmosphere. This layer causes electromagnetic propagation delays, such that incoming wavefronts are perturbed, which predominantly introduces phase errors (and in some cases amplitude errors). These perturbations vary in time, direction and frequency and also depend on the point where the electromagnetic field is measured on the ground. The errors are difficult to calibrate for due to their rapid time decorrelation. 
This section describes the effects of the ionosphere on radio-interferometric data. We focus on phase errors only and leave higher-order effects, such as differential Faraday rotation and the three-dimensional nature of the ionosphere, out of the scope of this work, because they are expected to affect the data to a much lesser extent \citep{gasperin_effect_2018}. 

\subsection{Frequency behaviour} \label{subsec:iono_freq}
The propagation delays, and resulting phase errors, are caused by the free electron density $N_e$ in the column of the ionosphere that the incoming wavefront traverses. This is quantified in the Total Electron Content (TEC), measured in TEC Units ($1\ \mathrm{TECU} = 10^{16}\ \mathrm{m^{-2}}$),
\begin{equation}
    \mathrm{TEC} \approx \int N_e (\mathbfit{s})\mathrm{d}s\ \mathrm{[m^{-2}]}. 
    \label{eq:iono_vTEC}
\end{equation}
In this equation, $\mathbfit{s}$ denotes the line of sight along which the TEC is measured. For LOFAR HBA frequencies, the phase delay $\Delta \phi$ in radians at frequency $\nu$ can be computed using
\begin{equation}
    \Delta\phi = \frac{e^2}{2\pi \nu c m_e \epsilon_\mathrm{0}} \mathrm{\left[\frac{TEC}{TECU}\right]}\approx 8.4\cdot 10^9 \cdot\mathrm{\left[\frac{TEC}{TECU}\right]}\frac{1}{\nu}\ .
    \label{eq:iono_phase_delay}
\end{equation}
 Here, $e$ is the electron charge, $\nu$ the frequency of the incident wave in Hz, $c$ the speed of light in vacuum, $m_e$ the rest mass of an electron and $\epsilon_\mathrm{0}$ the vacuum permittivity. The phase delay can be predicted across the entire waveband if the TEC is known. 
\begin{figure}
    \centering
    \includegraphics[width=\linewidth]{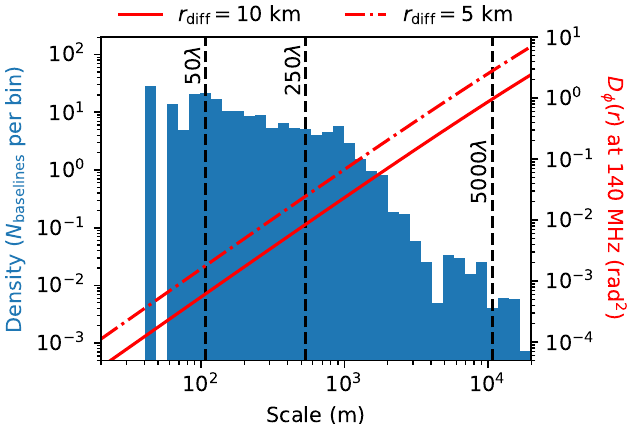}
    \caption{Projected baseline density of a LOFAR observation (left axis) and phase structure function (right axis) for both a diffractive scale of 5 and 10~km. The baseline lengths are binned in 40 logarithmically spaced bins between 20~m and 20~km. The baseline lengths of $50$, $250$ and $5000~\lambda$ are also indicated for a frequency of 140~MHz.}
    \label{fig:iono_bldensDf}
\end{figure}

\subsection{Spatiotemporal behaviour}\label{subsec:iono_spacetime}
In addition to its spectral variation, the TEC varies spatially over a range of length scales and in time. In this paper, we use the frozen screen approximation, where the differential TEC that causes ionospheric turbulence is modelled as an unchanging screen that moves over the telescope \citep{taylor_spectrum_1938}. This movement causes the differential TEC observed by the stationary array to evolve with time.  The statistical properties of the phase errors are modelled by the phase structure function $D_{\phi}$, which describes the phase variance between two piercing points through the ionospheric screen as a function of the distance between them \citep{narayan_physics_1992,van_der_tol_bayesian_2009, vedantham_scintillation_2016}. A commonly used phase structure function is that of a Kolmogorov screen of the form
\begin{equation}
    D_{\phi}(r)=\left(\frac{r}{r_{\mathrm{diff}}}\right)^\beta\ .
    \label{eq:iono_Df}
\end{equation}
Here, $D_{\phi}$ is given in $\mathrm{rad}^2$, $r$ is the scale being probed and $r_{\mathrm{diff}}$ the diffractive scale, defined as the scale at which $D_{\phi}(r)=1~\mathrm{rad}^2$. We use 150~MHz as the frequency for which we define the diffractive scale\footnote{This frequency was chosen since it is the customary reference frequency for LOFAR EoR observations. Therefore, using 150~MHz makes it easier to compare our results to previous work, and we choose it despite it being outside of our simulation band (see \cref{tab:methods_MS}).}. A smaller diffractive scale means more phase variance on all scales (see the slanted lines in \cref{fig:iono_bldensDf}). A turbulent ionosphere is described by this equation using the Kolmogorov index $\beta$ of 5/3 \citep{rufenach_power-law_1972}.\par
Observations show that this structure function closely matches the ionospheric perturbations experienced by LOFAR \citep{2016RaSc...51..927M, gasperin_systematic_2019, gan_statistical_2022,waszewski_measurement_2022}. A notable difference is that the true index $\beta$ has been shown to vary somewhat, reaching levels of nearly 2 \citep{2016RaSc...51..927M}. This index indicates a more structured ionosphere, for example, due to Travelling Ionospheric Disturbances (TIDs) \citep{wandzura_meaning_1980,loi_density_2016}. Since the effects of large-scale TIDs are easier to calibrate for \citep{gasperin_effect_2018}, we choose to adopt the more commonly used value of the index of $\beta=5/3$. Some ionospheric models also include an inner and an outer scale in their phase structure functions, where the screen behaves differently from a pure Kolmogorov screen.  The inner scale is of the order of metres \citep{booker_role_1979,2015MNRAS.453..925V}. The outer scale at 150~MHz was observationally found to be at least $\sim$80~km \citep{2016RaSc...51..927M}. Like in real LOFAR EoR observations, we use baselines between 50~$\lambda$ and $5000$~$\lambda$ for calibration and power spectrum analysis, corresponding to physical lengths between $\sim$100~m and $\sim $11.2~km in the simulated frequency band. These baseline lengths are also indicated in \cref{fig:iono_bldensDf}. Since all of our baselines are sufficiently longer than the inner scale and shorter than the outer scale, these scales are omitted in the description used here. Finally, we do not include simulations of ionospheric ducts or other anisotropies, such as those discussed by \citet{trott_assessment_2018}.\par

\subsection{The role of baseline length in sensitivity to ionospheric errors} \label{subsec:iono_bl}
We use the thin-screen approximation and the pierce-point approximation to model ionospheric effects, because the thickness of the ionospheric layer is a second-order effect. The thin-screen approximation means that we do not model the full three-dimensional structure of the ionosphere, but instead simulate a two-dimensional screen. The ionosphere is simulated using the TEC at the pierce-points and scaled based on the angle under which incident radiation would travel through the ionospheric slab (see \citealt{edler_investigating_2021}). A pierce-point is defined as the point at which the line of sight of a station observing in a direction crosses the thin ionospheric screen. The ionospheric phase difference that a baseline experiences in a direction is given by the difference in TEC value between its pierce points. Under these assumptions, the pierce-point separation is the projected baseline length onto the phase screen. \par
According to the phase structure function in \cref{eq:iono_Df}, a longer baseline experiences a larger phase variance. The impact of the largest variations on those baselines is mitigated by the fact that long baselines can be better calibrated for ionospheric disturbances due to their large time coherence\footnote{These long baselines are also perturbed by the smaller scale variations between their pierce points, however.}.\par
\citet{lonsdale_configuration_2005} defined four ionospheric regimes, based on the size of the array and its Field of View (FoV) compared to ionospheric scales. LOFAR is in the regime where both its FoV and its longest baselines are large compared to the scales of the ionospheric irregularities \citep{van_der_tol_self-calibration_2007}. Its FoV means that the phase delay varies over the FoV, and therefore, direction-dependent calibration solutions are needed to correct for the ionosphere. Furthermore, because the stations are far apart geographically, such a set of direction-dependent solutions is needed for each station separately. \par
\citet{2015MNRAS.453..925V} show that for baselines shorter than the Fresnel scale $r_\mathrm{F}$, and for a source in the target direction, the decorrelation time $t_\mathrm{d}$ becomes $t_\mathrm{d}=2r_\mathrm{F}/\varv$, where $\varv$ is the bulk wind velocity. For baselines longer than the Fresnel scale, the decorrelation time of a baseline of physical length $b$ varies between $t_\mathrm{d}=2b/\varv$ and $t_\mathrm{d}=4b/\varv$, depending on the angle between the baseline and the wind direction\footnote{This holds for a frozen screen model. The decorrelation time is longer if the baseline is oriented in the direction of the motion of the ionosphere.}. We use a frozen screen moving at $100\ \mathrm{m~s^{-1}}$. For baselines shorter than the Fresnel scale of $\sim300$~m (which occurs for a thin-screen height of 300~km), this implies a decorrelation time of $\sim6~$s. The longest baselines used in this work are up to $\sim 11$~km (5000~$\lambda$). The decorrelation time of the largest ionospheric scales they are sensitive to can go up to a few minutes, but they are also influenced by the more rapidly decorrelating small-scale variations. Solution intervals during gain calibration, on the other hand, range from 30~s for initial calibration up to 20~min for direction-dependent calibration in the LOFAR EoR project (see \citealt{mertens_improved_2020}). Hence, rapid small-scale variations within a solution interval are not corrected for on these baselines, leading to phase errors on the data \citep{koopmans_ionospheric_2010}. This is especially true for the small-scale ionospheric variations, where the solution intervals far exceed the coherence time and hence, all ionospheric errors remain in the data after calibration. This also holds for any other instrument, such as MWA, HERA and SKA. Some `speckle noise', i.e. a halo-structure caused by the ionosphere around point-like sources, may also occur \citep{koopmans_ionospheric_2010}. This is often not seen in low-frequency images, because this speckle noise is strongly
baseline-dependent and often below the thermal or confusion noise. Furthermore, it is often dominated by the decorrelation of the Point Spread Function (PSF) of the array \citep{2015MNRAS.453..925V}. However, in 21\nobreakdash-cm experiments, it may show up as additional excess variance in the power spectrum due to its high sensitivity. These errors on the station-based gain solutions are not straightforward to predict and, furthermore, are strongly direction-dependent. Each station is part of baselines of various lengths and therefore experiences a range of wave-modes and coherence time scales simultaneously.\par

\subsection{Impact on the power spectrum}
\citet{gan_statistical_2022} have correlated observed excess variance with ionospheric activity parameters as measured over 3\nobreakdash-h time bins. No clear correlation between these parameters and the observed excess variance was shown. A stronger correlation between the excess variance and local sidereal time was found instead. However, this does not exclude the possibility that the ionosphere still plays a major role in the excess variance for two reasons. First of all, ionospheric errors make the removal of bright foregrounds more challenging. This implies that errors in foreground subtraction would also correlate with the on-sky position of these sources and therefore local sidereal time. This would mean an ionospheric effect may still show a predominant correlation with local time. Furthermore, the activity level of the ionosphere is difficult to measure with LOFAR, because short bursts of high ionospheric activity can dominate ionospheric variance estimates. A time bin may therefore have data that are only occasionally strongly affected by the ionosphere, whereas most data are only weakly affected by ionospheric errors. In this work, we assume that the ionospheric behaviour remains ergodic over time, such that there are no temporal changes in the phase-structure functions and therefore no ionospheric activity bursts. \par
The expected impact of ionospheric scintillation on the LOFAR 21\nobreakdash-cm  power spectrum has been estimated analytically by \citet{vedantham_scintillation_2016}. They have shown that for an instrument similar to LOFAR, in the case of a simplified sky power spectrum and calibration strategy, the excess variance above thermal noise follows a wedge-like structure, that is dependent on the ionospheric activity level. This structure leaves a part of the power spectrum (often called the `EoR window') mostly unaffected. However, a full analysis of a realistic sky and calibration pipeline was outside of the scope of their work. \citet{ewall-wice_impact_2017} have shown that modelling errors can cause foreground leakage into the EoR window. Since ionospheric errors cause distortions of foreground sources and therefore have a similar effect to modelling errors, this supports the idea that ionospheric errors may leak into other parts of the power spectrum. \par
Direction-dependent gain calibration in the LOFAR EoR pipeline (see \citealt{patil_upper_2017,mertens_improved_2020}) employs a calibration step where the full set of baselines is divided into a group of longer and a group of shorter baselines. The long baselines are used to calibrate gains. These gains are then applied to the sky model and this model is subtracted on shorter baselines. This avoids signal suppression on shorter baselines \citep{patil_systematic_2016,mevius_numerical_2022}, but may transfer gain-calibration errors due to the ionosphere from the longer to the shorter baselines, if the calibration intervals on the long baselines are longer than their coherence time. Whereas this transfer of gain errors has been investigated for incomplete sky models and beam errors \citep{barry_calibration_2016,ewall-wice_impact_2017,mourisardarabadi_quantifying_2019}, to our knowledge the impact of the transfer of ionospheric errors from longer to shorter baselines during calibration, and the impact of speckle noise, has not been investigated in detail via simulation but only estimated theoretically. \par

\section{Simulation and calibration pipelines} \label{sec:methods}
To gauge the effects of the ionosphere on LOFAR HBA data and data processing, we create realistic mock data sets via forward simulations, both including and excluding ionospheric errors, and calibrate those using a pipeline nearly identical to the state-of-the-art LOFAR EoR standard pipeline (Mertens et al., in preparation) that is described in \cref{app:ref_pipeline} (from here on referred to as the `standard pipeline'). The pipeline for simulation is described in \cref{subsec:methods_sim} and the pipeline for calibration and power spectrum estimation in \cref{subsec:methods_cal}. \par

\subsection{Simulation pipeline}\label{subsec:methods_sim}
\begin{figure*}
    \centering\includegraphics[scale=0.525]{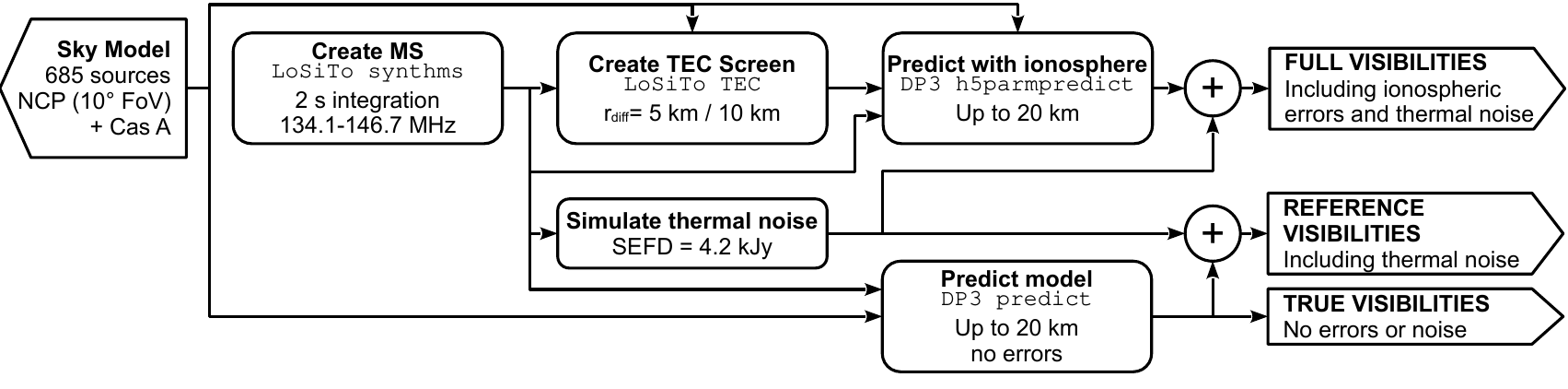}
    \caption{Pipeline used to create the simulated visibilities. The input sky model is denoted with a pentagon pointing leftwards. Processing steps are shown with a rounded rectangle. The title of the step is shown in bold font, the used software package is in mono-spaced font and any further details are noted in regular font. The circles with a `+' sign denote an addition in visibility space. The pentagons pointing to the right denote output simulated visibilities saved in the MS.}
    \label{fig:methods_simpipe}
\end{figure*}
To isolate the impact of the ionosphere from other sources of errors and systematics, our forward simulations contain a limited number of effects. The included components are a sky model, ionospheric phase-screen, the station beam models, and thermal noise contributions. The simulations are limited to 12~h, a full rotational synthesis. Because the 21\nobreakdash-cm signal is much weaker than the thermal noise level in observations of this duration, there is no need to include a 21\nobreakdash-cm signal in the simulations. The simulation process is described below and shown schematically in \cref{fig:methods_simpipe}. \par

\subsubsection{Sky model} \label{subsubsec:methods_sim_model}
\begin{figure}
    \centering
    \includegraphics[width=\linewidth]{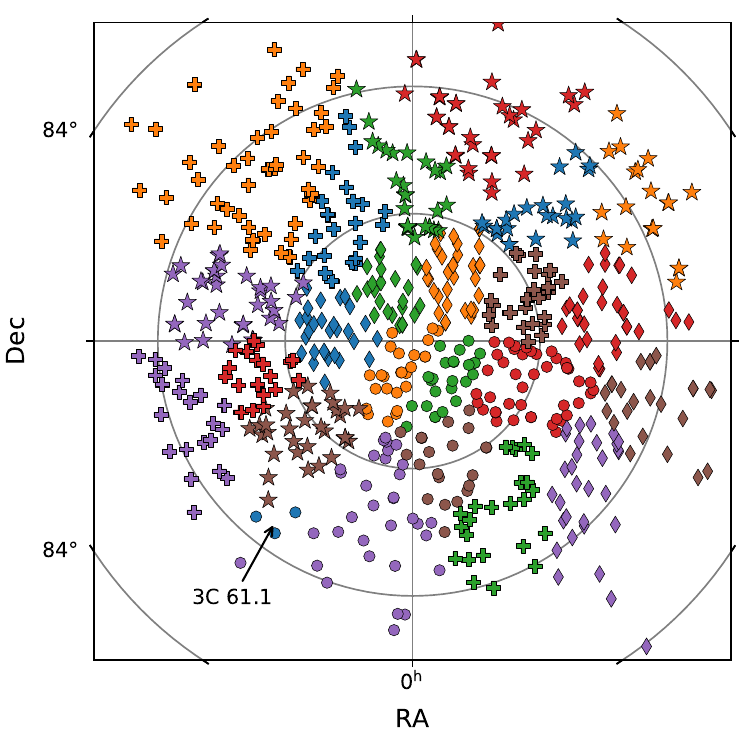}
    \caption{Schematic representation of the sky model used in this work. Each combination of shape and colour represents a calibration cluster for which a single gain solution is inferred and applied during sky-model subtraction. Due to its large angular separation from the NCP, Cas A is not shown in this figure, although it is included in the simulations.}
    \label{fig:methods_sim_model}
\end{figure}
The sky model is a representation of the North Celestial Pole (NCP) field at 145~MHz. The NCP field has been used to produce upper limits on the EoR signal with LOFAR \citep{patil_upper_2017,mertens_improved_2020} and NenuFAR \citep{munshi_first_2024}. \citet{2015MNRAS.453..925V} show that the square of the scintillation noise of a set of point sources is roughly equal to the sum of squares of the scintillation noises of the individual sources. As such, the scintillation noise due to all foregrounds can be well-approximated by only simulating the brightest sources. We limit the sky model to a restricted model of 684 unpolarised flat-spectrum sources, rather than using the 28,778 source components model used in the calibration of real LOFAR EoR data by \citet{mertens_improved_2020}. 
The restricted model was produced with  {\scriptsize WSCLEAN} \citep{offringa_wsclean_2014,offringa_optimized_2017}\footnote{W-Stacking CLEAN, \url{https://wsclean.readthedocs.io/en/latest/index.html}} on a small LOFAR dataset with a $10\times10^\circ$ FoV. Clean-components within the scale of the PSF were merged into single point sources. We set the intrinsic flux scale of the model using the predicted apparent flux of $\sim$100 sources. The restricted model and full model agree to a one per cent level in apparent flux. The cylindrically-averaged power spectra of the models agree to within ten per cent, sufficient to assess the impact of the ionosphere in a realistic scenario. \par
Because the spatial coherence scale of the ionosphere is roughly 3.5 arcmin at the frequencies of interest on the baselines used to create the power spectra \citep{vedantham_scintillation_2016}, one would ideally compute a gain solution for each patch of this spatial scale on the sky. This is not possible, however, because this would create too many degrees of freedom during calibration and therefore increase the risk of 21\nobreakdash-cm signal suppression. Furthermore, the signal\nobreakdash-to\nobreakdash-noise ratios of the patches would become too low to reach accurate gain solutions \citep{kazemi_clustered_2013}. Instead, the 684~sources in our model are divided into only 24 clusters for gain calibration. This is a factor of 4-5 fewer than used in the standard pipeline, because we do not solve for the instrument beam during calibration and because the coherence time of the ionosphere on the power-spectrum baselines is much smaller than the calibration time interval. The bright source 3C~61.1 is located near the first beam null in the NCP field. The apparent brightness of this source varies strongly, because the station beam rotates around the NCP. Due to its distinctive role in calibration strategy (see \cref{subsubsec:methods_cal_DD}), we assign 3C~61.1 to its own cluster\footnote{This was done based on source position alone, such that the actual cluster consists of a few sources as can be seen in \cref{fig:methods_sim_model}.} and divide the remaining sources into 23 clusters based on their positions and apparent flux using Voronoi tesselation with {\scriptsize{LSMTOOL}}\footnote{LOFAR Sky Model Tool, \url{https://lsmtool.readthedocs.io/en/latest}}. Our clustered model is shown in \cref{fig:methods_sim_model} and has been included as supplementary material to this paper. An excerpt of the model is shown in \cref{app:sky model}. \par
Due to its extreme apparent brightness, the bright far-field source Cassiopeia A (Cas A) has been included in the simulations\footnote{Using the LOFAR Low-Band Array model taken from \url{https://github.com/lofar-astron/prefactor/tree/master/skymodels}}. While Cygnus A (Cyg A) is also taken into consideration when the standard pipeline is applied to real data, it has a peak apparent brightness that is an order of magnitude lower than Cas A in the simulated LST range. Its impact on the power spectrum, which scales with the flux squared, is therefore much lower. See, for example, \citet{cook_investigating_2022}, who show the impact of unsubtracted far-field sources on the power spectrum. In their figures 6 and 7, they show the removal of a single bright source, the effect of which dominates over other, weaker off-axis sources through most of the power spectrum. In a similar way, Cas A dominates over the far-field foregrounds, and we use it as a representative source to describe the general behaviour of bright far-field sources on the power spectrum and omit Cyg A.

\subsubsection{Measurement set} \label{subsubsec:methods_sim_MS}
Measurement sets (MS) are created using the LOFAR Simulation Tool ({\scriptsize LOSITO})\footnote{\url{https://losito.readthedocs.io}}. An MS contains properties of the telescope array, such as station positions, $u\varv$-coverage and frequency channels. The MS is adjusted to a specified phase centre (target direction) and observation time. We simulate the Dutch HBA stations used in the EoR project \citep{patil_upper_2017,mertens_improved_2020} with a realistic pattern of inactive tiles (groups of 4$\times$4 antennas) within the station. Two types of stations are included: Core Stations (CS) that are concentrated within a few kilometres diameter area near the LOFAR core, and Remote Stations (RS) that predominantly provide longer baselines. Remote stations have twice the number of tiles that core stations have and are beamformed to closely match the station layout of core stations by assigning zero weights to a sub-set of the tiles. In real data, baselines sharing a common electronics cabinet lead to crosstalk and are therefore removed, so they are removed in the simulation as well.\par
A time resolution of 2~s is used. This is the highest time resolution data stored for LOFAR EoR observations and captures temporal variations at the same scale as available in measured data. Furthermore, this time interval is shorter than the temporal coherence at the Fresnel scale of $\sim$6~s \citep{2015MNRAS.453..925V}, such that it is smaller than the shortest ionospheric timescales. The frequency resolution is set to 195~kHz, the same spacing as the smallest spectral gain calibration intervals used in real data. As described in \cref{eq:iono_phase_delay}, the spectral behaviour of the ionosphere is smooth, such that a higher spectral resolution is not needed for this simulation. Because we aim to assess the impact of the ionosphere, there is no need to simulate spectral effects introduced by the receiver chain of the telescope. The total simulated bandwidth is 12.6~MHz, between 134.1~MHz and 146.7~MHz (the same as used for $z=9.1$ observations). We apply a baseline filter removing baselines of $>20~$km as they are much longer than the longest baselines used during data processing ($\sim 11$~km). Due to the distribution of stations, this leaves our longest baseline at $18~$km, reducing the size of the TEC-screen needed.

\begin{table}
    \caption{Properties of the simulated measurement sets}
    \label{tab:methods_MS}
    \begin{tabular}{lc}
        \hline
        Parameter & Value \\
        \hline
        Telescope & LOFAR HBA\\
        Pointing & RA = $0^\mathrm{h}00^\mathrm{m}00^\mathrm{s}$, Dec = $+90^{\circ}00'00''$\\
        Sky model & 684 sources in the central $10\times 10^\circ$\\
        & Cassiopeia A\\
        Bandwidth & 134.1--146.7~MHz\\
        Frequency resolution & 195~kHz\\
        Time resolution & 2~s\\
        Total duration & 12~hr\\
        System equivalent flux density & 4.157 kJy\\
        Baseline range & 42~m -- 18~km\\
        Height of ionospheric layer & 300~km\\
        \hline
    \end{tabular}
\end{table}

\subsubsection{TEC Screen} \label{subsubsec:methods_sim_TEC}
We use the TEC method in {\scriptsize LOSITO} as described in \citet{edler_investigating_2021} to model the ionosphere. A frozen two-dimensional TEC screen is created for a given diffractive scale at 150~MHz at a given height (300~km in this work). This is done using a combination of a large-scale and small-scale screen as proposed by \citet{buscher_simulating_2016}, such that both scales can be computed in an efficient manner. Subsequently, pierce-points are computed for each source and station separately. The TEC values at these pierce-points are found through interpolation. A geometric factor is applied to the TEC value at each pierce-point to correct for the angle at which the ionosphere is pierced. The geometric factor takes into account both the angle of incidence and the curvature of the Earth. Hence, lines of sight near the horizon experience a higher TEC and therefore smaller diffractive scale than points near the zenith. This can lead to bright off-axis sources dominating the ionospheric errors when they are low in the sky as shown by \citet{gehlot_wide-field_2018} and is only mitigated by beam attenuation.

\subsubsection{Thermal noise} \label{subsubsec:methods_sim_thermal}
Thermal noise is characterised by the System Equivalent Flux Density (SEFD). The used SEFD is 4.2~kJy, based on the estimated SEFD for LOFAR EoR observations listed by \citet{mertens_improved_2020}. Thermal noise is generated by drawing the real and imaginary component of the noise component of the visibility separately from a Gaussian distribution with a standard deviation of
\begin{equation}
    \sigma_{\mathrm{thermal}} = \frac{\mathrm{SEFD}}{\sqrt{2 \Delta t\Delta\nu}}\ \mathrm{[Jy]}.
\end{equation} 
Here, $\Delta t$ and $\Delta\nu$ represent the temporal and spectral resolution of the visibilities in the MS as listed in \cref{tab:methods_MS}.\par
The thermal noise is stored separately, such that the same noise realisation can be added to several simulated datasets. This allows for a better comparison between calibration results for different simulations while still including the effect of noise.

\subsubsection{Visibility prediction} \label{subsubsec:methods_sim_predict}
Forward prediction of the visibilities is done using {\scriptsize DP3} \citep{van_diepen_dppp_2018}\footnote{Default Pre-Processing Pipeline}. {\scriptsize LOSITO} has been designed to interface with {\scriptsize DP3}, such that the TEC \nobreakdash screen can easily be applied in a direction-dependent way during the prediction step. {\scriptsize DP3} allows for control over the simulation process, such that we can choose not to include temporal and spectral smearing in our simulations and therefore use a lower spectral resolution for the simulations (see \cref{tab:methods_MS}).
{\scriptsize DP3} includes a realistic model of the LOFAR beam including sidelobes. This is important for simulating the behaviour of Cas A, which can pass through such sidelobes during the observations. We produce three data products:

\begin{itemize}
    \item\texttt{TRUE VISIBILITIES} -- This set of visibilities is a forward prediction of the sky model with LOFAR beam effects and contains no noise or ionospheric effects. It is considered the `ideal case' and can be compared to other data products to quantify their ionospheric errors. 
    \item\texttt{REFERENCE VISIBILITIES} -- This set is created by adding a thermal noise realisation to the \texttt{TRUE VISIBILITIES}. We use this set to assess the impact of an ionosphere-free to a full model, which includes ionospheric effects
    \item \texttt{FULL VISIBILITIES} -- This set is generated by adding phase perturbations to the predicted visibilities of each source separately using the TEC\nobreakdash-screen described in \cref{subsubsec:methods_sim_TEC} to create direction-dependent ionospheric effects. Furthermore, the same thermal noise realisation is added as to the \texttt{REFERENCE VISIBILITIES}.
\end{itemize}

\subsection{Calibration and power spectrum estimation pipeline} \label{subsec:methods_cal}

\begin{figure*}
    \centering
    \includegraphics[width=\textwidth]{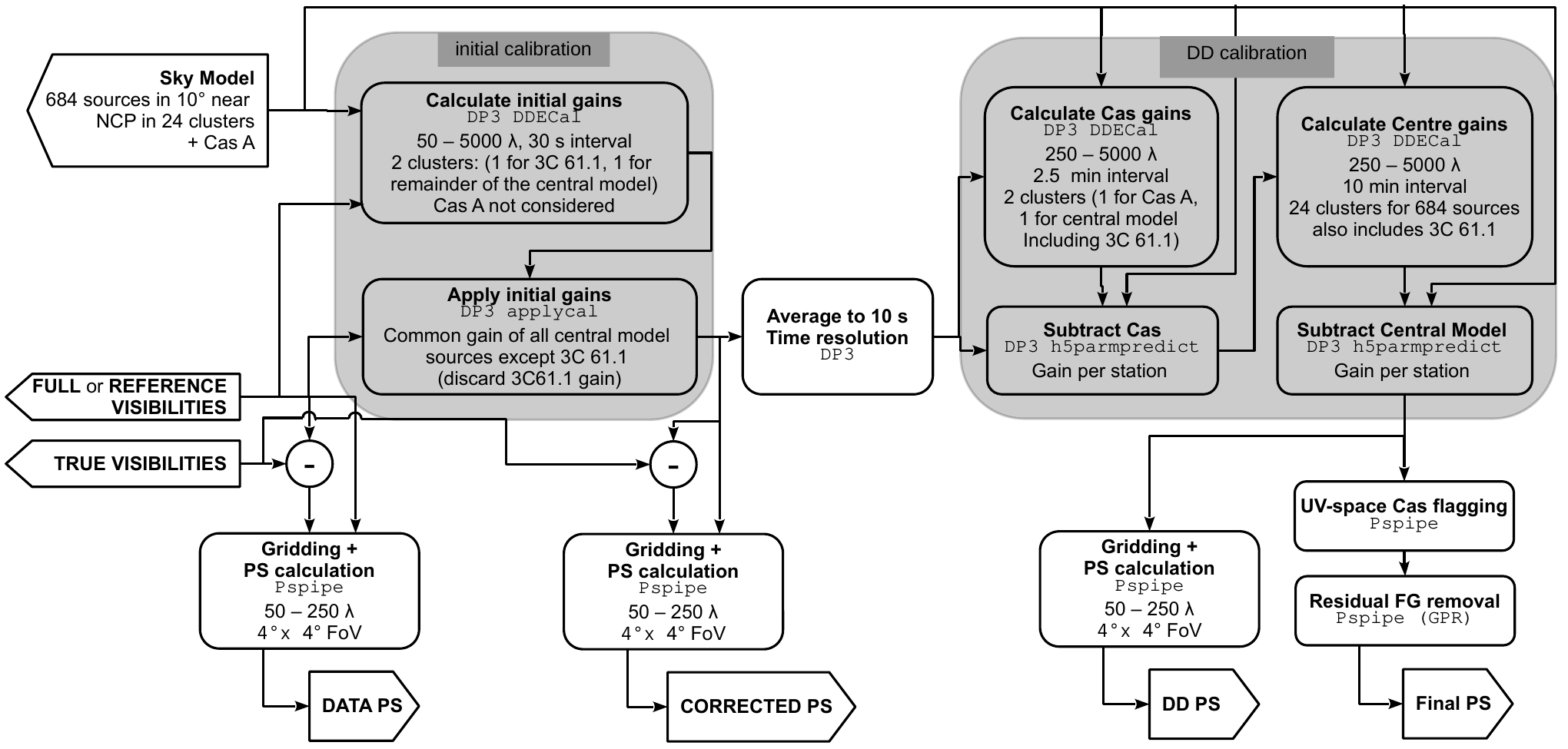}
    \caption{Pipeline used to calibrate the simulated visibilities. The input visibilities simulated using the pipeline shown in \cref{fig:methods_simpipe} are denoted with a pentagon pointing to the left. Processing steps are shown with a rounded rectangle. The title of the step is shown in bold font, the used software package in mono-spaced font and any further details are noted in regular font. The circles with a `-' sign denote a subtraction in visibility space. The pentagons pointing to the right denote output power spectra shown in \cref{sec:results}.}
    \label{fig:methods_calpipe}
\end{figure*}
The calibration and power spectrum estimation pipeline largely follows the LOFAR EoR standard pipeline (Mertens et al., in preparation, see \cref{app:ref_pipeline}). Because our input data is simplified compared to observational LOFAR EoR data (such as the omission of a 21\nobreakdash-cm signal, bandpass errors, RFI, etc.), some parts of the standard pipeline can be omitted, resulting in the pipeline shown in \cref{fig:methods_calpipe}. However, there are a few differences between this pipeline and the standard pipeline:

\begin{itemize}
    \item The calibration pipeline presented in this paper utilises the {\scriptsize DDECal} \citep{gan_assessing_2023} submodule of {\scriptsize DP3} instead of {\scriptsize SAGECAL-CO} \citep{yatawatta_distributed_2015}. As mentioned in \cref{subsubsec:methods_sim_predict}, there are advantages to using {\scriptsize DP3} for simulating the data. However, because {\scriptsize SAGECAL-CO} and {\scriptsize DP3} use slightly different beam models, we prefer to use {\scriptsize DP3} in both simulation and calibration, as it allows us to use the exact same beam model in both. Therefore, any ambiguity between beam modelling mismatches and ionospheric effects is avoided. As shown by \citet{gan_assessing_2023}, calibration pipelines using {\scriptsize DDECal} and  {\scriptsize SAGECAL-CO} produce comparable power spectra, such that the impact of this difference in calibration algorithm will minimally affect the resulting power spectra. Additionally, differential experiments have shown that the difference in calibration algorithm is a second-order effect in the calibration result. The differences in the calibration results between {\scriptsize SAGECAL-CO} and {\scriptsize DDECal} are a factor $10^{6-7}$ smaller than the source powers (Gehlot, private communications).
    \item Assigning solution intervals that differ between source clusters is difficult, because of technical differences between {\scriptsize DDECal} and {\scriptsize SAGECAL-CO}\footnote{This is because {\scriptsize DDECAL} performs gain calibration by iterating over stations rather than calibration directions. While {\scriptsize DDECal} does offer the functionality of direction-dependent calibration intervals, it comes at the cost of sacrificing precision and robustness by treating all directions independently within the update step.}. Therefore, we use the two-step method presented by \citet{gan_assessing_2023}, where bright off-axis sources are subtracted at a higher time resolution in a separate step. This step is followed by a lower time resolution subtraction of the weaker sources near the target direction. 
    \item Radio frequency interference flagging and bandpass calibration steps are omitted, because these effects are not present in the simulated data. 
    \item We only consider a single night of observation at a time for power spectrum estimation.
\end{itemize}

\begin{table*}
    \caption{Settings for initial and Direction-Dependent (DD) calibration}
    \label{tab:methods_calsettings}
    \begin{threeparttable}
        \begin{tabular}{lccc}
            \hline
             \multirow{2}{*}{Property} & \multirow{2}{*}{Value in Initial calibration} & Value in DD-calibration&Value in DD-calibration\\
             &&for Cas A &  for near phase-centre sources\\
             \hline
             Operation & Apply gains & Subtract with gains& Subtract with gains\\
             Minimum baseline $(\lambda)$& 50 & 250& 250\\
             Maximum baseline $(\lambda)$& 5000 & 5000& 5000\\
             Number of directions & 2\tnote{a} & 2\tnote{b} & 24\\
             Solution time interval (s) &  30 & 150&600\\
             Solution frequency interval (kHz) & 195 & 195 & 195\\
             Number of iterations & 100 & 200& 200\\
             Jones matrix elements calibrated & Diagonal & Diagonal& Diagonal \\
             Smoothness Constraint & 4~MHz & 4~MHz & 4~MHz\\
             \hline
        \end{tabular}
        \begin{tablenotes}
            \item[a] 3C~61.1 is placed in one cluster, the remainder of the central model is placed in the other.
            \item[b] Cas A is placed in one cluster, the full central model in the other.
        \end{tablenotes}
    \end{threeparttable}
\end{table*}

\subsubsection{Initial calibration} \label{subsubsec:methods_cal_DI}
The settings in all gain calibration steps are listed in \cref{tab:methods_calsettings}. 
In the first gain-calibration step, a common gain across the field is computed for each station. In the case of the NCP field, this round of calibration is complicated by the bright source 3C~61.1 near the first beam-null, which can cause time- and direction-dependent effects during calibration. Initial calibration is, therefore, done in two directions, such that 3C~61.1 can be isolated. The gains computed in the direction of 3C~61.1 are discarded, and the gain solutions computed for the rest of the field are applied to the full dataset. Cas A is not included in any of the initial calibration, because its apparent brightness is highly temporally variable due to beam sidelobes far from the target direction.\par
Initial calibration is performed on a solution interval of 30~s, for baselines ranging from $50$ to $5000~\lambda$. We use the spectral smoothness constraint, with a width of the Gaussian smoothing kernel of 4~MHz recommended by \citet{gan_assessing_2023}. Because the sky model is unpolarised, instrumental polarisation errors are not introduced, and dispersive effects of the ionosphere produce a scalar gain perturbation\footnote{Higher order effects such as differential Faraday rotation are not simulated and therefore do not need to be corrected for.}. Only gain solutions on the diagonal of the Jones matrices are computed.

\subsubsection{Direction-dependent calibration} \label{subsubsec:methods_cal_DD}
The direction-dependent calibration step removes the sources in the sky model from the data. Gains are computed in the directions of the sources, and then the contributions to the visibilities from each source are predicted and subtracted using those gains.
Before Direction-Dependent (DD) gain calibration, the data are first averaged to a time resolution of 10~s. Only the longer baselines of length $>250~\lambda$ are used during this calibration step to avoid subtraction of the 21\nobreakdash-cm signal of interest during the calibration process \citep{patil_systematic_2016}. \par
We first subtract Cas A in a separate calibration step with a time resolution of 2.5 minutes. To do this, the sky model is divided into two clusters: a cluster for Cas A and a cluster containing all other sources. After calibration, the gains computed for Cas A are used to subtract it from the data, whereas the other gains are discarded.
Subsequently, a second DD-calibration step is performed to remove the remaining 684~sources near the target direction. We perform DD-calibration towards 24 clusters of sources on a 10-minute time resolution. This process is summarised in the right grey box in \cref{fig:methods_calpipe}. 

\subsubsection{A-team flagging} \label{subsubsec:methods_cal_casflag}
\begin{figure}
    \centering
    \includegraphics[width=\linewidth]{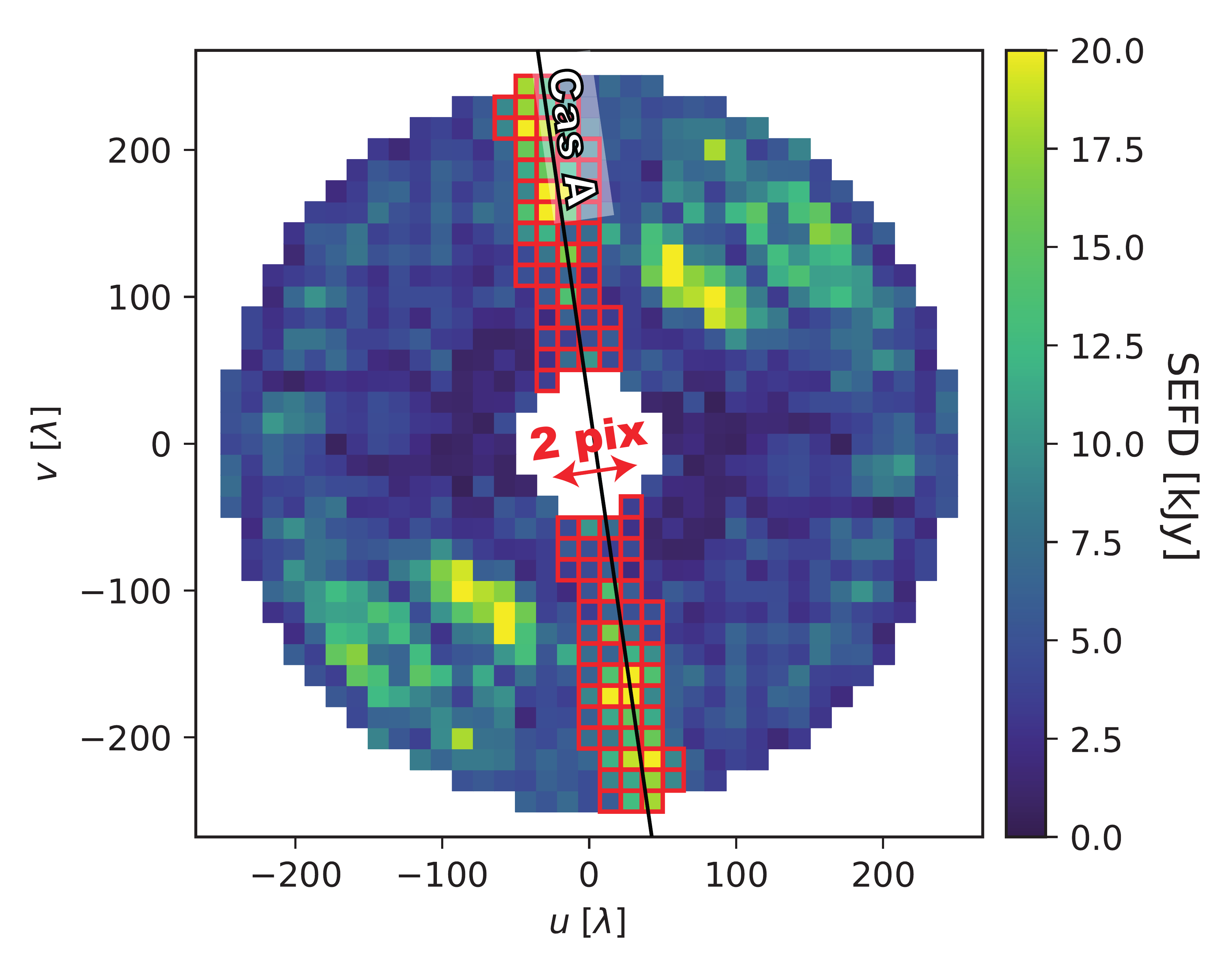}
    \caption{Illustration of $u\varv$-space masking. The coloured pixels show the SEFD estimated from the spectrally differenced simulated visibilities of the central model and Cas A. A line in the direction of Cas A is shown in black and the pixels highlighted in red represent the pixels discarded during ($u,\varv$)-space flagging.}
    \label{fig:methods_uvmask}
\end{figure}

After DD gain-calibration and sky model subtraction, the dynamic range in the data is drastically reduced and the data can be gridded in $u\varv$-space. Because the baselines migrate through $u\varv$-space as a function of time and frequency and multiple baselines are gridded into a single $u\varv$-cell, this step mixes residual post-calibration artefacts (e.g. from the ionosphere) between different baselines. Therefore, these artefacts can no longer be removed after gridding.
In real data, a few bright A-team sources still create large residuals in the power spectra at this point in the pipeline. They must be flagged before further analysis in a step called `$u\varv$-space masking' or `A-team flagging'. This is done by creating a mask in $u\varv$-space where the A-team sources are expected to be brightest (see \citealt{munshi_beyond_2024}). A line in the direction of each A-team source on the $u\varv$-plane is calculated, and all gridded visibilities within 2 pixels of this line are masked. This is illustrated in \cref{fig:methods_uvmask}. Although the A-team flagging step removes the strongest effects of these sources, not all of their impact is removed.

\subsubsection{Residual foreground removal} \label{subsubsec:methods_cal_GPR}
After DD-source subtraction, grinding, and A-team flagging, residual foregrounds can remain in the data. The spectral variability of the foregrounds is increased, due to a process called `mode-mixing' \citep{morales_four_2012}, where the chromatic nature of the interferometer mixes spatial fluctuations into the spectral domain, such that they obscure the much fainter 21\nobreakdash-cm signal. As a result, the foregrounds dominate over the 21\nobreakdash-cm signal in a larger region of the power spectrum than just the most spectrally smooth modes \citep{datta_bright_2010,vedantham_imaging_2012}. In addition, chromatic effects due to the ionosphere \citep{koopmans_ionospheric_2010,vedantham_scintillation_2016,trott_assessment_2018} and instrument beam can cause power from the foreground sources to leak to other modes. Because of this, these residual foregrounds need to be removed in order to recover the 21\nobreakdash-cm signal.\par
To remove the residual foregrounds from the data, \citet{mertens_statistical_2018} introduced a method based on Gaussian Process Regression (GPR). Gaussian processes are used to model the components in the data through fitting frequency covariances (kernels) as a function of $u\varv$-coordinates. These components can describe a specific process (such as the 21\nobreakdash-cm signal, mode-mixing, thermal noise, etc.), but also more general residual foregrounds or systematics (such as excess variance). The covariance kernels are described by a general functional form that depends on the frequency separation and several hyperparameters. These hyperparameters can describe e.g. the power normalisation or spectral coherence scale. Markov Chain Monte Carlo or Nested Sampling is used to obtain samples from the posterior probability distribution of the hyperparameters given the observed (or simulated) data. To compute a power spectrum of a component, a number of realisations are drawn from the posterior distribution for its kernel. Those realisations are used to compute a probability distribution of the power spectrum of the component. Its power spectrum is calculated from realisations drawn from this distribution.\par
In real data, the residual foregrounds are comprised of foreground sources not included in the model (such as diffuse emission or extragalactic emission near the confusion noise limit), and foreground sources that are not well subtracted (for example due to calibration errors). We expect to need fewer and slightly different components to model the data in our simulations than needed in real data. This is because our data contain no 21\nobreakdash-cm signal, beam errors nor sky-model incompleteness. Fewer foreground components may therefore be needed to describe the residual foregrounds after DD-subtraction. However, foreground contamination is still expected, due to ionospheric variations on timescales shorter than the calibration intervals which result in source subtraction errors.\par 
In this paper, we have therefore chosen a simplified set of kernels that subtract the foregrounds to a satisfactory level. This set might not be optimal nor unique but demonstrates the behaviour of ionospheric errors in GPR.
The used GPR components are the intrinsic foregrounds, mode-mixing contaminants, and uncorrelated thermal noise. The intrinsic foreground kernel is intended to capture smooth emission that has not been accurately removed during sky-model subtraction inside the image field used for gridding. Similar to \citet{mertens_improved_2020}, we use a Median Radial Basis Function (MRBF) with a large length scale (150~MHz, much larger than the total bandwidth of the simulations) to model this emission. The mode-mixing kernel describes spectrally rapidly fluctuating effects introduced by the instrument, such as the imprint of the PSF and beam chromaticity on residual foregrounds both inside and outside the gridding image. The latter is of a lesser concern in this work, due to the limited angular extent of our sky model. Mode-mixing is modelled with a 5/2 Mat\'ern covariance kernel, which is narrower than the MRBF kernel. Finally, thermal noise is modelled to be uncorrelated in frequency. \par
A noise estimate is made through the Stokes V temporal differences in the visibilities. The noise variance is scaled with respect to this estimate. All other variances are normalised to the data variance. The hyperparameters are listed in \cref{tab:methods_gprsettings}. A nested sampler with 500 live points is used to optimise the evidence for each simulation. \par

\begin{table*}
    \caption{List of used hyperparameters in GPR. In this table $\mathcal{U}$ and $\mathcal{LU}$ indicate a uniform and log-uniform prior respectively. This is a simplified set of kernels in comparison to the kernels used in the standard pipeline, because many effects present in real data have been omitted from our simulations to isolate ionospheric effects. More details on these kernels can be found in \citet{mertens_statistical_2018,mertens_retrieving_2024}.}
    \label{tab:methods_gprsettings}
    \centering
    \begin{tabular}{lccc}
        \hline
        {Component} & {Kernel} & Hyperparameter & {Prior} \\
        \hline
        \multirow{2}{*}{{Intrinsic foregrounds}} & \multirow{2}{*}{MRBF}&{Data normalised variance}& {$\mathcal{LU}(10^{-4},10^2)$}\\
        &&Length scale & Fixed at 150\\[.2cm]
        \multirow{2}{*}{{Mode-mixing foreground}}&\multirow{2}{*}{{5/2 Mat\'ern kernel}}&Data normalised variance&$\mathcal{LU}(10^{-4},10^{2})$\\
        &&Length scale&$\mathcal{U}(0.1,20)$\\[.2cm]
        Noise&Uncorrelated&Scaling with respect to Stokes V time difference&$\mathcal{U}(0.75,1.3)$\\
        \hline
    \end{tabular}
\end{table*}

\subsubsection{Power spectra} \label{subsubsec:methods_cal_PS}
Due to the extreme sensitivity needed to detect the 21\nobreakdash-cm signal, several modes are averaged in Fourier space in order to create a suitably sensitive power spectrum. A version of this power spectrum often used to gauge the effectiveness of data analysis pipelines is the cylindrically averaged power spectrum. In this spectrum, the Fourier dual of the frequency channels of an observation are used on one axis ($k_\parallel$). The other axis ($k_\perp$) averages over the Fourier dual of spatial variations on the sky. As such, the pixels of the cylindrically averaged power spectrum represent annuli in Fourier space. Ideally, the spectrally smooth foregrounds would be confined to the lowest $k_\parallel$-modes, leaving the remainder of the power spectrum free for 21\nobreakdash-cm signal analysis. However, the aforementioned effect of `mode-mixing' causes the foregrounds to dominate over the 21\nobreakdash-cm signal in a larger wedge-shaped region of the power spectrum, called the `foreground-wedge'. \par
The power spectra in this work are computed using the code {\scriptsize PSPIPE}\footnote{Power-Spectrum PIPEline, \url{https://gitlab.com/flomertens/pspipe}} with the conventional definitions by \citet{morales_toward_2004}. Only the shortest baselines (50--250~$\lambda$) which are the most sensitive to the 21\nobreakdash-cm signal are used for this. A power spectrum is produced at each calibration step, to investigate how the ionosphere impacts that step. \par
Assessing errors made during the DD-calibration step can be difficult, because the DD-calibration step also includes sky-model subtraction. This means that, even if calibration errors are introduced, the foreground power is typically reduced by orders of magnitude. Therefore, we have differenced the \texttt{FULL VISIBILITIES} and \texttt{TRUE VISIBILITIES}, both before and after initial calibration of the \texttt{FULL VISIBILITIES}, and created power spectra of the residual as well. Although this can not be done with real data, it provides a fairer comparison of the ionospheric effects on the data.

\section{Results and Discussion}\label{sec:results} 
In this section, we analyse the effects of the ionosphere on the simulations described in \cref{subsec:results_simulations}. We first inspect how ionospheric errors appear in the cylindrically averaged power spectrum in different steps of the calibration and data processing pipeline in \cref{subsec:results_steps}. The impact of the level of ionospheric turbulence (i.e. different diffractive scales) is compared in \cref{subsec:results_5vs10}. A deeper inspection of the effects A-team flagging is performed in \cref{subsec:result_casflag}. The effects of phase and amplitude calibration respectively are analysed in \cref{subsubsec:results_steps_partial}. Finally, the correlation between nights is inspected in \cref{subsec:results_timecorr}, to investigate whether ionospheric errors average down when multiple nights of data are integrated.

\subsection{Simulations} \label{subsec:results_simulations}
Five simulations created using the simulation pipeline described \cref{subsec:methods_sim} are summarised in \cref{tab:results_observationoverview}. The first two simulations (\texttt{r10a} and \texttt{r10b}) assume a diffractive scale of 10~km, and are the most representative of LOFAR EoR data (see \citealt{2016RaSc...51..927M, gan_statistical_2022}). These simulations are used to investigate the coherence of ionospheric residuals in observations at different local sidereal times (LST). A different LST means that the baseline orientations and station beams compared to the sky will be different. However, there may still be some coherence between the Fourier modes probed within a gridded $u\varv$-bin across multiple nights, such that the residuals of strong sources remain correlated. As such, these simulations make it possible to estimate whether the residuals are incoherent across multiple nights and can therefore be suppressed through processing more data.\par
Two simulations represent an active ionosphere and have a diffractive scale of 5~km. This scale is the limit at which LOFAR EoR data is considered to be of sufficient quality. One (\texttt{r5b}) of these two simulations has a unique TEC screen, and one (\texttt{r5a}) has the TEC screen used in one of the 10~km diffractive scale simulations (\texttt{r10a}), with all TEC values scaled up to a diffractive scale of 5~km. This allows us to differentiate between effects due to stronger ionospheric activity for the exact same screen and between different random realisations.\par
Finally, we have performed a simulation that does include thermal noise, but no ionospheric errors (\texttt{Null}). This enables differentiation between effects due to the ionosphere and effects due to data processing, calibration (e.g. gain-solution time scale), sky-model subtraction and GPR (e.g. chosen kernels). All simulations that share a start time also share the same thermal noise realisation. This minimises ambiguity in whether different results in the various simulations can be attributed to the ionospheric or thermal noise realisations. To create a realistic second `night' of simulation, the simulation for another start time (\texttt{r10b}) has been given a unique thermal noise realisation.\par

\begin{table*}
    \caption{Overview of the simulated data used in this work. A diffractive scale of 5~km is the cut-off for real data, below which data is not used for processing due to its high distortion level.}
    \label{tab:results_observationoverview}
    \centering
    \begin{tabular}{lcccc}
         \hline 
         Label & $r_\mathrm{diff}$ & Start time (UTC) & TEC realisation & Noise realisation\\
         \hline 
         \texttt{r10a}& 10~km & 2013-11-27 18:00:00 & Nominal model & Nominal model \\
         \texttt{r10b} & 10~km & 2014-03-06 18:00:00 & Unique & Unique\\
         \texttt{r5a} &  5~km  & 2013-11-27 18:00:00 & Scaled TEC-values of \texttt{r10a}& Same as \texttt{r10a}\\
         \texttt{r5b} &5~km  & 2013-11-27 18:00:00 & Unique & Same as \texttt{r10a}\\
         \texttt{Null}& $\infty$ &2013-11-27 18:00:00 & No ionospheric errors & Same as \texttt{r10a}\\
         \hline
    \end{tabular}
\end{table*}

\subsection{The propagation of ionospheric errors}\label{subsec:results_steps}
To assess how ionospheric errors in the calibration process evolve, first, we analyse the errors in the nominal simulation \texttt{r10a} separately and compare it to the ionosphere-free case in the \texttt{Null} simulation.

\begin{figure*}
    \centering
    \includegraphics[width=\textwidth]{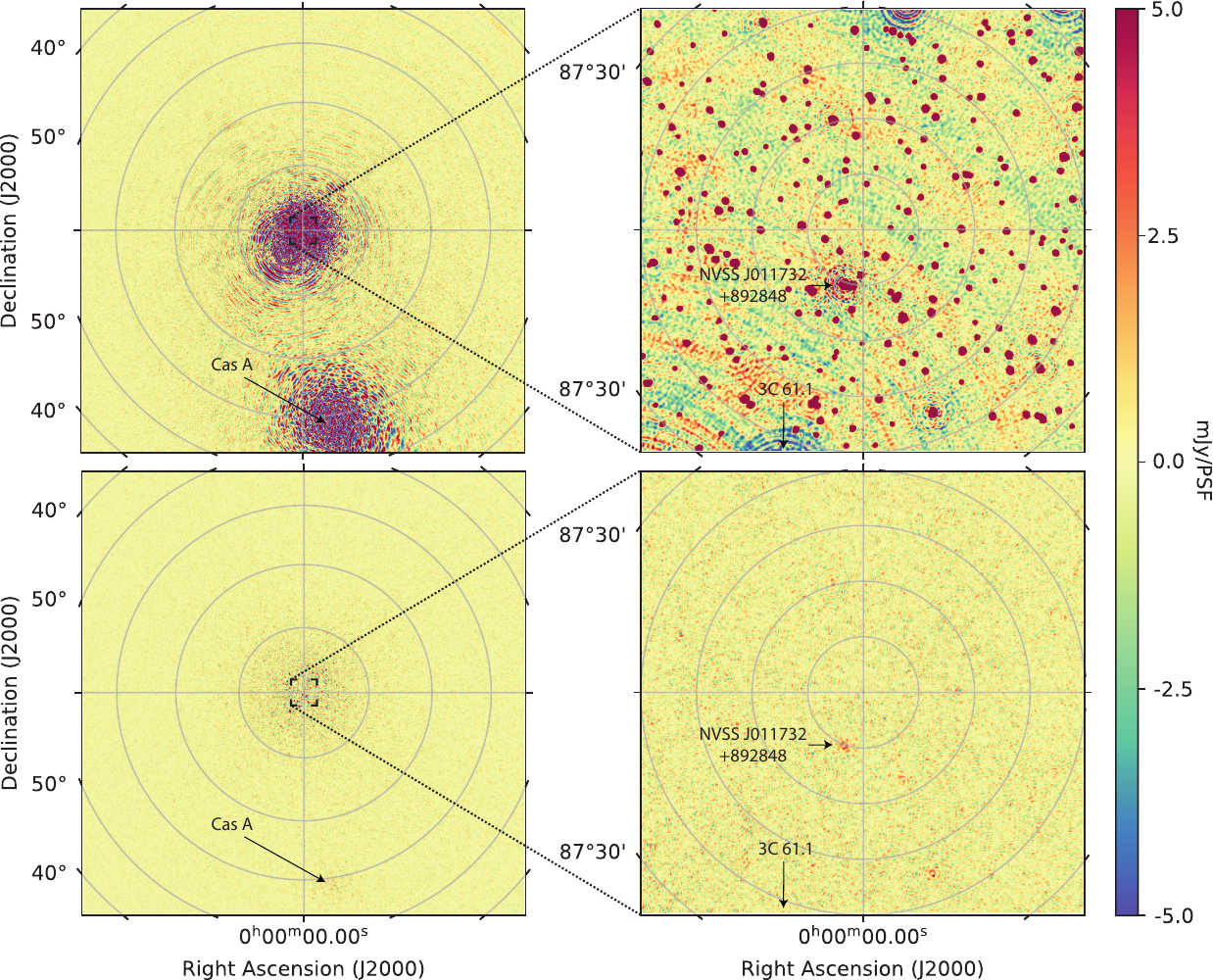}
    \caption{Images created using the full nominal simulation \texttt{r10a} with {\scriptsize WSCLEAN}. The top panels are created with initial-calibrated data and the bottom panels with DD-source subtracted data. The left panels represent a $68^{\circ}\times68^{\circ}$ field of view with a 7~arcmin resolution made using baselines between 50 and 1000~$\lambda$. The right panels are a zoomed-in version showing only the central $4^{\circ}\times4^{\circ}$, with a 42~arcsec resolution and baselines between 50 and 5000~$\lambda$. A few important sources are labelled. Power near the NCP and around Cas A is clearly visible in the initial-calibrated wide field image (top left). The flux of both Cas A and the 684~source model is reduced after DD-calibration, although residual power is still visible.}
    \label{fig:results_images}
\end{figure*}

\begin{figure*}
    \centering
    \includegraphics[width=\textwidth]{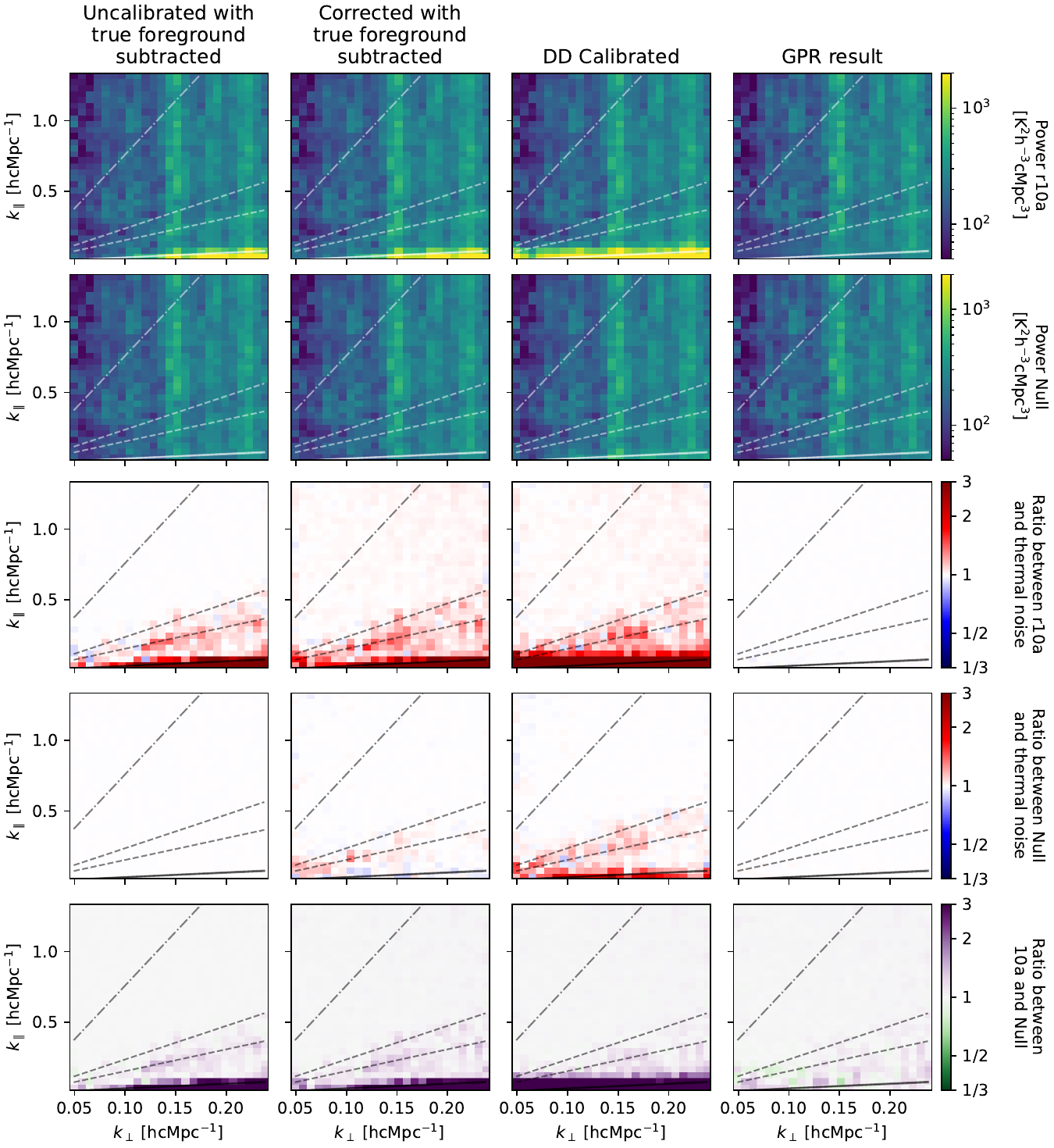}
    \caption{Overview of the residual power in cylindrical power spectra of nominal simulation \texttt{r10a} and the ionosphere-free simulation \texttt{Null}. The columns represent the different calibration steps. From left to right: before any calibration, after initial calibration, after direction-dependent calibration with source subtraction and after Gaussian process regression. The \texttt{TRUE VISIBILITIES} have been subtracted from the data in the first two columns, such that all columns represent residuals, even before the source subtraction step in the pipeline. From top to bottom, the rows represent the power spectra for \texttt{r10a} and \texttt{Null} respectively, the ratio between these power spectra and a thermal noise power spectrum, and the ratio between the power spectrum of simulation \texttt{r10a} and that of simulation \texttt{Null}. In the first three columns of the middle row, the thermal noise is the injected thermal noise. In the last column, this is replaced by a thermal noise cube drawn from GPR (as described in \cref{subsubsec:methods_cal_GPR}). In this cylindrically averaged power spectrum and in all such power spectra throughout this work we plot four delay lines, from top to bottom: the horizon line, two lines defining the delay area where the peak power from Cas A should lie, and the delay below which the peak power of a source at 5$^\circ$ from the phase centre should lie. These lines are based on improved delay calculations that take the sky curvature and phase-referencing away from zenith into account by \citet{munshi_beyond_2024}.}
    \label{fig:results_overzichtsplot10km}
\end{figure*} 

\begin{table*}
    \centering
    \caption{Mean and maximum power spectrum ratios between simulations and the thermal noise or \texttt{Null} simulation respectively. The shown simulations all share the same LST as the \texttt{Null} simulation. The lowest level subcolumn represents the point in the pipeline; after initial calibration, after DD-source subtraction and after GPR respectively. For the ratio between GPR and thermal noise, two ratios are shown: the first is the ratio with the noise cube drawn from GPR (the same as used in the noise ratios in the fourth column of \cref{fig:results_overzichtsplot10km}), the second is the ratio with the input thermal noise realisation (as in the other thermal noise ratios). This second value suffers from sampling variance after GPR, making the maximum more prone to outliers. We therefore refer to the first ratio when discussing excess variance, and provide the second one for completeness.}
    \label{tab:results_ratios}
    \begin{tabular}{l@{\hskip 0.5cm}cccc@{\hskip 0.9cm}cccc@{\hskip 0.9cm}ccc@{\hskip 0.9cm}ccc}
    \hline
    \multirow{5}{*}{Label} & \multicolumn{8}{c}{Ratio with thermal noise} & \multicolumn{6}{c}{Ratio with \texttt{Null} simulation} \\
    \cline{2-9}
    \cline{9-15}
    & \multicolumn{4}{c}{Maximum} {\hskip 0.9cm} & \multicolumn{4}{c}{Mean} {\hskip 0.9cm}& \multicolumn{3}{c}{Maximum} {\hskip 0.9cm}& \multicolumn{3}{c}{Mean} \\
    & \multirow{3}{*}{Initial} & \multirow{3}{*}{DD} & GPR&{GPR}  & \multirow{3}{*}{Initial} & \multirow{3}{*}{DD} & GPR&{GPR}  & \multirow{3}{*}{Initial} & \multirow{3}{*}{DD} & \multirow{3}{*}{GPR} & \multirow{3}{*}{Initial} & \multirow{3}{*}{DD} & \multirow{3}{*}{GPR}  \\
    &&&versus&versus&&&versus&versus\\
    &&&estimate&input&&&estimate&input\\
    \hline
    \texttt{r10a}& 12.6&22.3&1.0&1.9&    1.2&1.6&1.0&1.0&     13.0&13.2&1.4&     1.2&1.4&1.0\\
    \texttt{r5a}&  39.8&454.5&1.1&2.7&   1.7&9.7&1.0&1.1&     41.2&237.3&2.6&    1.7&6.4&1.1\\
    \texttt{r5b}&  30.7&801.2&1.0&2.1&   1.7&16.9&1.0&1.1&    32.5&558.1&1.8&    1.7&11.3&1.1\\
    \hline
    \end{tabular}
\end{table*}

\begin{figure*}
    \centering
    \includegraphics[width=\textwidth]{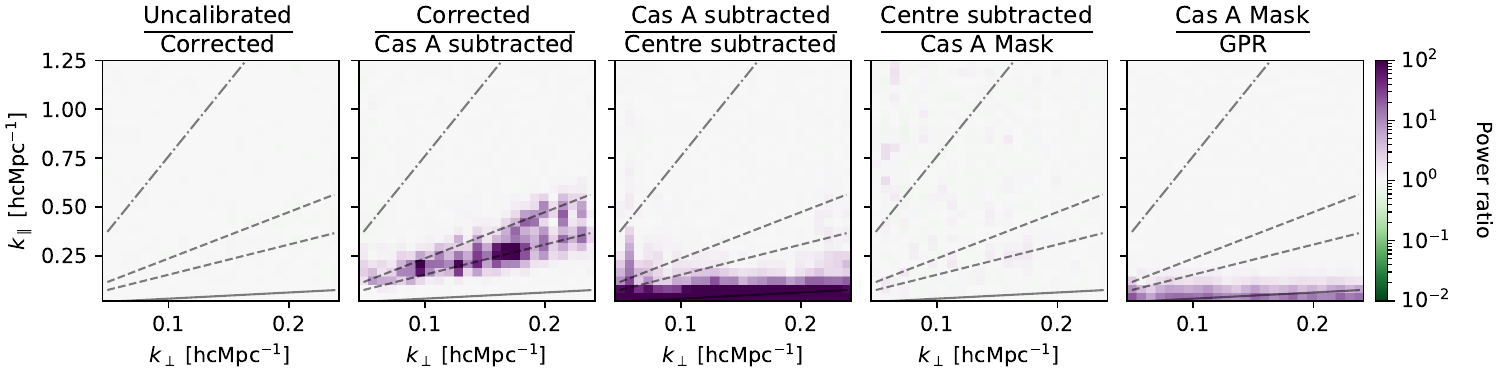}
    \caption{The ratio between cylindrically averaged power spectra between calibration steps for simulation \texttt{r10a} to illustrate the impact of each calibration step. From left to right, the columns represent: the ratio between raw data and initial-calibrated data, the ratio between initial-calibrated data and the same data after subtraction of Cas A, the ratio between the data after Cas A has been subtracted and after the central sky model has been subtracted, the ratio between this residual and the same data with Cas A masked out in A-team flagging, and the ratio between this data product and the same after residual foreground removal. }
    \label{fig:results_steps10km}
\end{figure*}

\subsubsection{Simulated data and initial calibration}
\cref{fig:results_images} shows (residual) images made using simulation \texttt{r10a}. The top row shows images after initial calibration. The bright source near the first null, 3C~61.1, causes the ring structures coming from the bottom-left corner in the initial-calibrated target field image. The overall behaviour of the simulated data in these images resembles that of real data\footnote{See for example the right side of Fig. 2 in \citet{mertens_improved_2020}, which is plotted at the same angular scale as our figure.}. \par
\cref{fig:results_overzichtsplot10km} shows the residual power spectra created at different stages of the calibration process for simulations \texttt{r10a} and \texttt{Null} (we present the same figure for the other simulations in \cref{app:overviews}). Note that the power spectra shown here are all residual power spectra, i.e. the \texttt{TRUE\ VISIBILITIES} have been subtracted from the raw and initial-calibrated data, in order to show the ionospheric errors rather than the source power. The ratio between the power spectrum of the simulations and their injected noise is presented in the third and fourth rows. They show the `excess' variance as any residual power above the thermal noise level. The bottom row, which shows the ratio between the power spectrum of simulations \texttt{r10a} and \texttt{Null}, discriminates between systematic errors due to the ionosphere, and any other possible systematic errors.\par
Up to the delay line of Cas A (the source farthest away from the phase centre), the residuals follow the wedge and are typically a few times the thermal noise (third row, panels one and two). We expect the residuals to be coherent as function of frequency, such that they can be removed in the GPR step. There is a slight excess in the ratio with thermal noise after initial-calibration above the wedge (the red hue starting just above {Cas A} and extending into the EoR window in \cref{fig:results_overzichtsplot10km} column 2 panel 3). This is due to the application of the gains from initial calibration, which change the visibilities and therefore also this region of the power spectrum. The ratio then changes, because the thermal noise we are comparing to is not calibrated.\par
In \cref{fig:results_steps10km}, changes in the cylindrically averaged power spectrum throughout the calibration pipeline are presented. We show the ratio of power spectra extracted at each subsequent stage of the pipeline. This figure highlights where power is added or removed and therefore, we use the \texttt{FULL VISIBILITIES} directly for these plots (rather than creating a residual like in \cref{fig:results_overzichtsplot10km}). In the leftmost panel, the ratio between un-calibrated and initial-calibrated data is shown. This ratio does not appreciably deviate from unity anywhere in the power spectrum. Large deviations would mainly occur for large phase variations that are coherent across the full FoV (tip-tilt corrections). However, the ionospheric phase variations in the simulation are small, because only baselines up to 250~$\lambda$ are used in the power spectra. 
Therefore, the changes between the raw and initial-calibrated data being minimal indicates that no major systematic errors or corrections are introduced at this point in the pipeline.

\subsubsection{Direction-dependent calibration}
The Cas A model is subtracted in the next step of the pipeline, followed by the NCP source model. The residuals in the image plane can be seen in the bottom two panels of \cref{fig:results_images}. While DD-calibration can remove most of the source flux, some power remains visible near {Cas A} in the wide field image. However, the impact of these residuals on the power spectrum is best estimated in the central $4\times4^\circ$ field of view, because this is where the power spectrum is estimated from. When comparing the variance of this part of the image to an image made with the same simulation, but with Cas A omitted from both the simulation and calibration, we find a $\sim0.5$ per cent change in variance in the full-band image and a $\sim0.1$ per cent change in variance in a single-channel image. We conclude that the impact of Cas A is limited on images in this FoV. In the target field image (bottom-right panel), the brightest sources leave a ring-like structure behind. This could either be an artefact of the PSF convolved with the residual of the source, or due to source scintillation creating a halo-like structure around the original source position directly \citep{koopmans_ionospheric_2010, vedantham_scintillation_2016}.\par
The effect of removing {Cas A} and the NCP sources on the power spectra is shown separately in \cref{fig:results_steps10km}. During the {Cas A}-subtraction step, power is removed mostly around the delay region of {Cas A}, with a slight extension above it at low $k_\perp$. Removal of the NCP sources results in power reduction at low $k_\parallel$, extending above the expected delay of the central sky model. This cannot be due to the spectral dependence of ionospheric phase errors on the short baselines directly, as such power should be visible in \cref{fig:results_steps10km}. Therefore, we attribute this to the transfer of errors from longer baselines to shorter ones, which introduces stronger spectral variations in the source power, extending the foreground power to higher $k_\parallel$-modes. Any transfer of errors happens along constant delay lines up to the maximum delay causing a so-called "brick" in the power spectrum \citep{ewall-wice_impact_2017}. Additionally, there is a vertical feature at $k_\perp=0.06$ in the third panel of \cref{fig:results_steps10km} (A similar feature is present in \cref{fig:app_overview_5km1,fig:app_overview_5km2}), attributed to low $u\varv$-coverage in this part of the spectrum. Baselines at this separation have been removed because $k_\perp=0.06~h\mathrm{cMpc^{-1}}$ corresponds to the separation between a pair of stations that share an electronics cabinet (see \citealt{ewall-wice_impact_2017}, Section 3.2). There is a gap in the $u\varv$-coverage around this $k_\perp$, because we remove these baselines in the simulation, similar to how we treat real data.\par
Additional power introduced during the DD-calibration steps are more clearly seen comparing the initial-calibrated and DD-source subtracted residuals in \cref{fig:results_overzichtsplot10km}. Because the \texttt{TRUE VISIBILITIES} have been subtracted from the visibility data shown in the second column of this figure, the change between these columns represents errors introduced by imperfect gain calibration during the DD-step. Note that this is not the same as a comparison between gains that perfectly compensate for the ionosphere and gains that do not, because the \texttt{TRUE VISIBILITIES} ignore the ionosphere and assume gains of unity. Power in the wedge is more pronounced in the power spectra of the residuals after DD-calibration (compared to the idealised foreground subtraction in the previous column), both at the delays of the NCP source and at the delay lines of {Cas A}. In the power spectrum itself (the first panel in the third column of \cref{fig:results_overzichtsplot10km}), this is mainly visible at the lowest $k$-modes and for {Cas A} around $k_\parallel = 0.2~h\mathrm{cMpc^{-1}}$.
When inspecting the ratio with thermal noise in the middle panel in the third column of \cref{fig:results_overzichtsplot10km}, the excess of Cas A is strongly pronounced. Interestingly, however, the excess power near the delay lines of {Cas A} is weak in the ratio between simulations \texttt{Null} and \texttt{r10a} (bottom row). The same effect is clear in \cref{tab:results_ratios}, where the mean and maximum values of the ratio between \texttt{r10a} and its thermal noise is higher than the ratio between \texttt{r10a} and \texttt{Null}. Because the residuals of Cas A are stronger if there are ionospheric errors, but do not disappear if no ionospheric errors are present, we conclude that  the increase in excess power at this stage in the data processing is linked to ionospheric errors, but can not be completely attributed to them.\par

\subsubsection{A-team flagging and residual foreground removal} \label{subssubec:results_steps_casflag} 
\begin{figure}
    \centering
    \includegraphics[width=\linewidth]{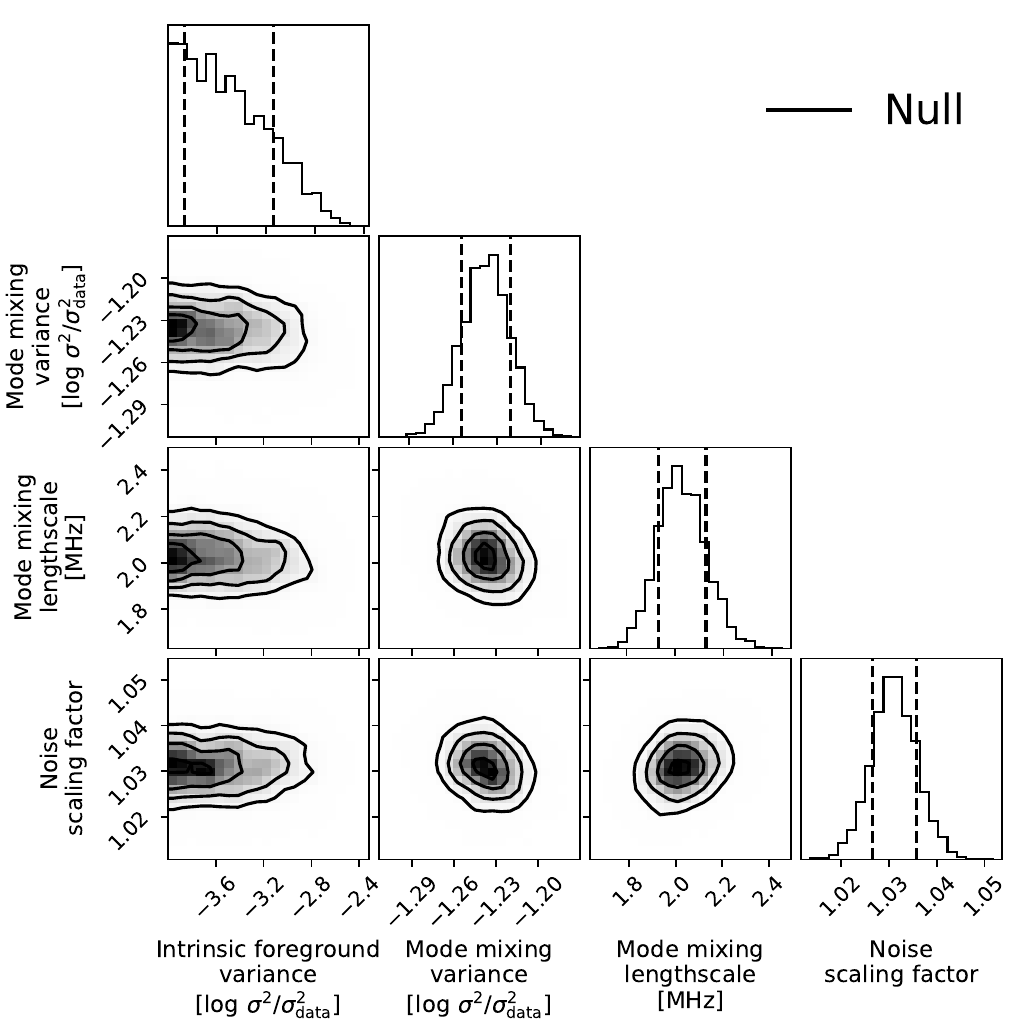}
    \caption{Corner plot of the posterior probability distributions of GPR on the \texttt{Null} simulation. This figure illustrates how the components are fitted in GPR. The vertical dashed lines in the marginalised distributions indicate the 68 per cent confidence level.}
    \label{fig:results_cornernull}
    \includegraphics[width=\linewidth]{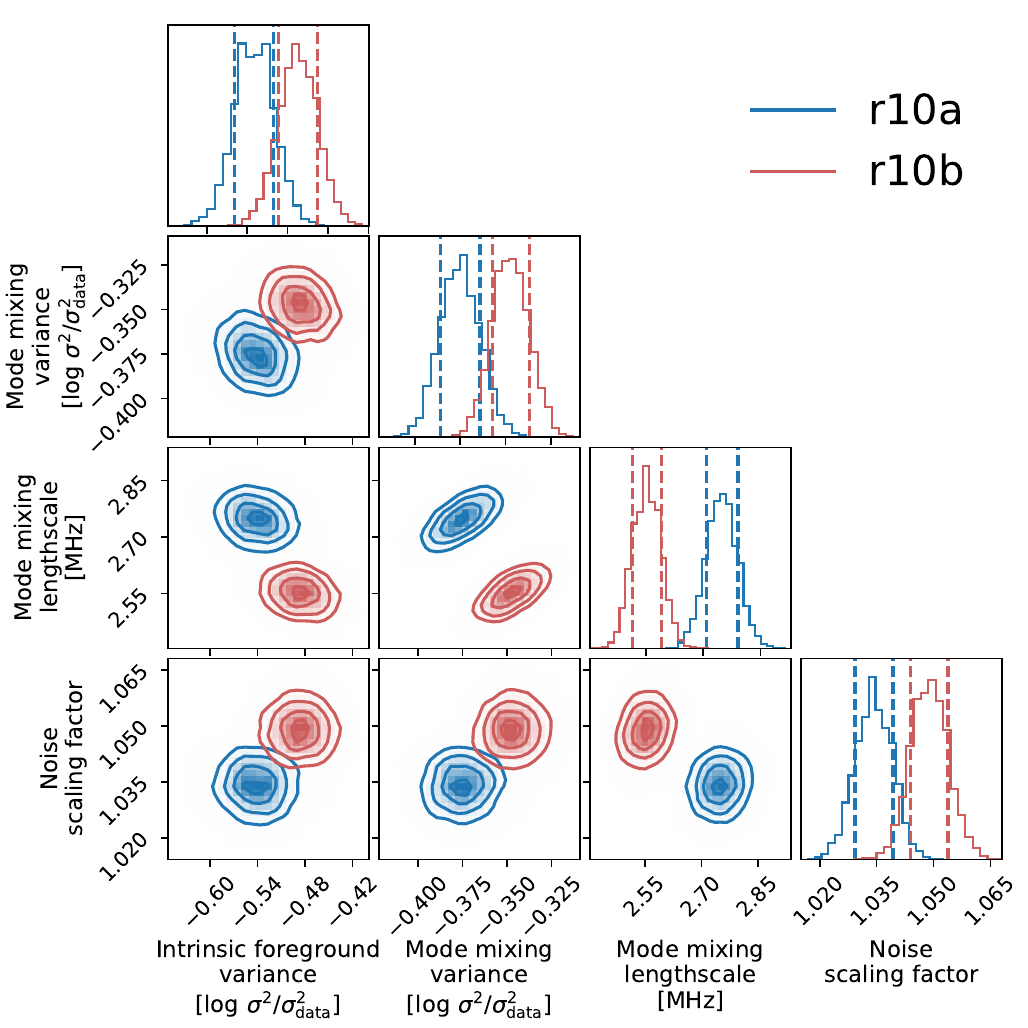}
    \caption{The same as \cref{fig:results_cornernull}, but with simulation \texttt{r10a} shown in blue and simulation \texttt{r10b} shown in red. Despite having the same ionospheric activity level, the fitted kernels are slightly different between simulations \texttt{r10a} and \texttt{r10b} due to different ionospheric and thermal noise realisations and due to the LST difference between the simulations.}
    \label{fig:results_corner10}
\end{figure}
\begin{figure}
    \centering
    \includegraphics[width=\linewidth]{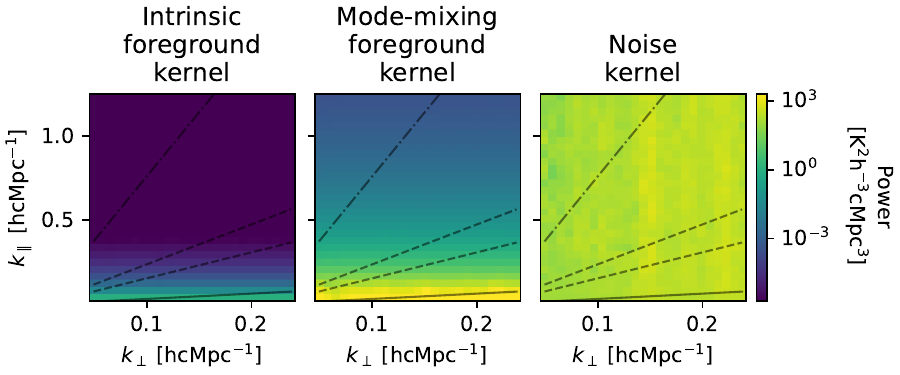}
    \caption{Cylindrically averaged power spectrum representation of the fitted covariance kernels produced by drawing 1,000 power spectra from each component to illustrate where they capture power. These realisations were generated using the hyperparameters found for simulation \texttt{r10a}.}
    \label{fig:results_GPRrealisations}
\end{figure}

During the final data processing steps,  the $u\varv$-cells where bright off-axis sources are expected to be strongest are masked and GPR model-fitting is applied to remove residual foregrounds present in the data after calibration and model subtraction. We show the posterior distributions of the GPR hyperparameters in \cref{fig:results_cornernull,fig:results_corner10} (the posteriors of the remaining distributions are shown in \cref{app:corner}). The cylindrically averaged power-spectra of the different components fitted for simulation \texttt{r10a} are shown in \cref{fig:results_GPRrealisations}.\par
The masking of {Cas A} only has a very minor effect in the fourth panel of \cref{fig:results_steps10km}, indicating that the source has been removed well during direction-dependent calibration. 
GPR has removed power near the target direction, as is evident in the lowest $k_\parallel$-modes. This is in line with expectations from the foreground kernels visualised in \cref{fig:results_GPRrealisations}.  
Most notably, we do not see an appreciable excess variance due to ionospheric errors in the final power spectrum of simulation \texttt{r10a}. Residual foreground removal effectively removes ionospheric errors from the cylindrically averaged power spectra to the thermal noise level. This is also evident from \cref{fig:results_overzichtsplot10km} and \cref{tab:results_ratios}, where no clear excess over thermal noise is observed anymore after GPR. There are some minor effects visible in the wedge region in the comparison between simulations \texttt{r10a} and \texttt{Null}. However, these are small and randomly distributed around unity, possibly speckle noise, indicating an increased variance in the power spectrum estimate, rather than a systematic bias. \par
Recall that the noise scaling hyperparameter is defined compared to the noise estimate from Stokes V time difference, and we therefore expect the noise power to be unity. However, there is a positive bias in noise scaling of the posteriors on the level of a few per cent. This issue persists when the priors are chosen differently, and when all variances are estimated in linear space rather than log-space. Therefore, the bias indicates that there is another power component being absorbed into the noise kernel. Because there is no clear systematic difference between the noise scaling in the \texttt{Null} simulation and the other 4 simulations (that do contain ionospheric effects), we do not expect this bias to be an ionospheric effect. We do not investigate it further here, because the resulting error is well below the level where it would significantly influence results in the analysis of real 21\nobreakdash-cm data, and because the bias does not appear to result from the ionosphere.\par
When comparing the GPR components, thermal noise is dominant outside of the lowest $k_\parallel$-modes. The foreground residuals are mostly captured by the mode-mixing kernel, but also to a lower level by the intrinsic foreground kernel. This is the decomposition of the ionospherically corrupted residuals in a spectrally smooth and more rapidly fluctuating component. That these are ionospheric effects is supported by the results for the \texttt{Null} simulation, where the variance of the intrinsic foreground kernel hits the lower bound of its prior and the mode-mixing component is much weaker than for the simulations affected by the ionosphere. The absence of ionospheric errors makes source subtraction more effective, hence fewer residual foregrounds are present before GPR. The spectral coherence scales found for the mode-mixing kernel are low: around or below 3~MHz. \citet{mertens_statistical_2018} use  3~MHz as the limit below which some signal loss is expected, if the 21\nobreakdash-cm signal itself is either not a part of the GPR model or incorporated in an approximate manner. The newly developed ML-GPR method proposed by \citet{mertens_retrieving_2024} and tested by \citet{acharya_21-cm_2024} has an improved kernel for modelling the 21\nobreakdash-cm component, such that it can be distinguished from spectrally varying foreground power more easily and the 3~MHz cut-off has become less stringent. As such, the low spectral coherence scale may not lead to 21\nobreakdash-cm signal subtraction in real data.\par
Although the current set of kernels is able to remove the ionospheric residuals to the noise level, two kernels are required to fully fit the ionospheric signature. For real data, this may pose a challenge in the presence of other errors, which also need to be removed by these kernels. Furthermore, the mode-mixing spectral coherence scale drops to a low spectral width, which could potentially lead to some signal loss. These issues may be remedied by creating a more physically motivated kernel for either instrument-induced mode-mixing or ionospheric effects.\par
The residual power spectra clearly differ from those obtained using LOFAR observations, as described by \citet{mertens_improved_2020}, both in shape and amplitude. For a similar length of observation and after similar data processing, the foreground wedge is still clearly visible in real data, whereas it can be removed in our simulations. This is despite real data having a similar or larger diffractive scale than the simulations, such that ionospheric effects should be smaller in real data. Furthermore, observed data also shows an increase in power for the lowest $k_\perp$-modes, which we do not reproduce in our simulations. We therefore exclude ionospheric errors as the dominant cause of the excess variance found in LOFAR EoR observations under normal ionospheric conditions of $r_{\mathrm{diff}}= 10~$km.\par

\subsection{Ionospheric conditions}\label{subsec:results_5vs10}
\begin{figure*}
    \centering
    \includegraphics[width=\textwidth]{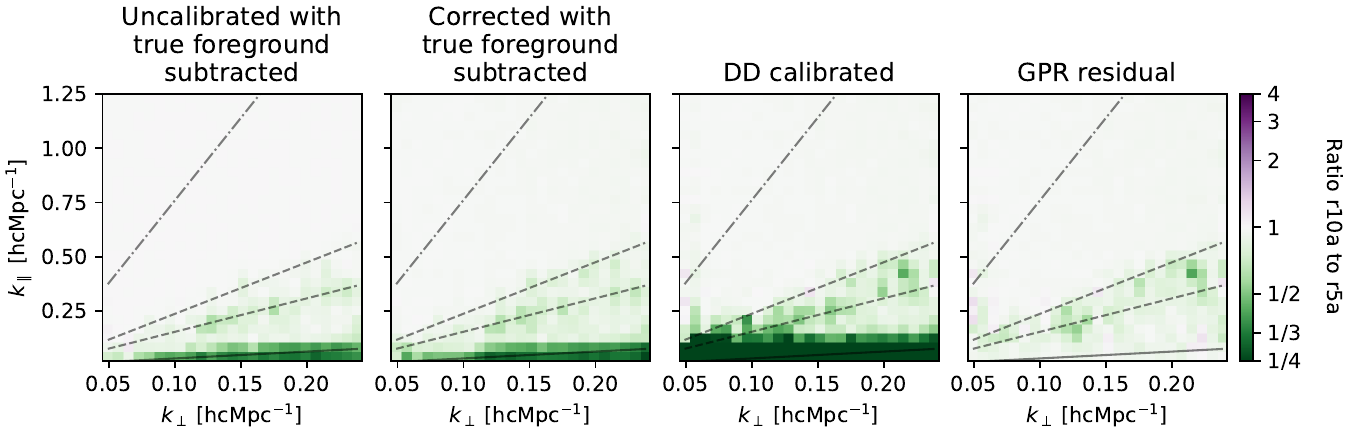}
    \caption{The ratio between cylindrically averaged residuals (computed in the same way as in \cref{fig:results_overzichtsplot10km}) of simulations \texttt{r10a} and \texttt{r5a}. These simulations share the same ionospheric phase screen, barring a re-scaling of all TEC values, such that the diffractive scale of simulation \texttt{r5a} simulation is half of that of simulation \texttt{r10a}. Their ratio illustrates the impact of the activity level of the ionosphere on each step of data processing. A green hue indicates more residual power in the simulation with a larger ionospheric variance. Note that outside the wedge, little to no difference is observed between the two power spectra as they share a thermal noise realisation.}
    \label{fig:results_stepscomp5and10}
\end{figure*}
Because simulations \texttt{r10a} and \texttt{r5a} are created using the same thermal noise realisation and TEC screen, be it scaled in amplitude in the latter case, the impact of the level of ionospheric distortions can be assessed without being affected by differences in the noise and ionospheric sample variance between the two simulations. A diffractive scale of 5~km  leads to an increase in residual power during calibration for two reasons. First of all, a smaller diffractive scale means more ionospheric phase errors on the short baselines on time scales shorter than the solution time interval for the gains, which is directly reflected in the residuals in the power spectrum. In conjunction with this, larger ionospheric variations on large spatial scales reduce the calibratability of long baselines and are therefore detrimental to the gain solutions which are applied to the shorter baselines as well. However, even for a diffractive scale of 5~km, complete decorrelation is not reached on the majority of baselines, because they are below the diffractive scale\footnote{The longest baselines do experience a higher level of decorrelation, because both the NCP sources and Cas A are positioned away from the zenith for at least part of the observation. Due to the increased air mass away from the zenith, the experienced diffractive scale is lower than that set for the TEC screen.}.\par
In \cref{fig:results_stepscomp5and10}, we present the ratios of the power spectra of \texttt{r10a} and \texttt{r5a} at different steps in the calibration pipeline. In the raw and initial-calibrated data, the effect of the larger ionospheric errors is visible in the foreground wedge, most prominently in the lower $k_\parallel$-modes. The difference between the two datasets becomes more pronounced after the DD-calibration step, as is expected due to the calibration limitations outlined above. The excess power at this point in the pipeline is about an order of magnitude higher for simulation \texttt{r5a} compared to simulation \texttt{r10a}, as is shown in \cref{tab:results_ratios}. This table also shows the same ratios for simulation \texttt{r5b}, which has a unique ionospheric TEC-screen. Although the evolution of excess variance of \texttt{r5a} and \texttt{r5b} through the pipeline follows a similar trend and the excess is clearly higher than in \texttt{r10a}, their excess variances are significantly different. We attribute the difference to sampling variance, but do conclude that a lower diffractive scale leads to larger foreground residuals. This demonstrates that the impact of the ionospheric activity level is a non-linear effect that can drastically affect the gains, motivating the need for an ionospheric activity cut-off. \par
In the right-most column of \cref{fig:results_stepscomp5and10}, slightly more residual power is present near the delay line of {Cas A} for the dataset with a more active ionosphere. This is expected to be a result of the non-optimal (i.e. hard flagging) $u\varv$-plane masking of {Cas A}. The $u\varv$-plane has been masked in the same way in both data sets, but because we only mask in a 2\nobreakdash-pixel width around the direction of Cas A, whereas its true footprint exists throughout the $u\varv$-plane, some residuals of Cas A can remain in the data after $(u,\varv)$-space flagging. Furthermore,  because GPR only removes power at the delay lines near the target direction rather than in the area of {Cas A}, it is not able to remove the residual of Cas A and a larger residual before $(u,\varv)$-space flagging also results in a larger residual after. This is true for both $r_\mathrm{diff}=5~$km simulations (as can be seen in the third panels of the fourth columns in  \cref{fig:app_overview_5km1,fig:app_overview_5km2}). However, this residual has very low power and is only visible here because the simulations have an identical thermal noise realisation. In real data, we do not probe down to the level of this residual within a single night, which can be seen from \cref{tab:results_ratios}. Therefore, based on these results ionospheric errors can even be removed in the worst ionospheric conditions used for data processing ($r_\mathrm{diff}=5~$km), with excess of at most 10 per cent.\par

\subsection{Necessity of A-team flagging}\label{subsec:result_casflag}
\begin{figure*}
    \centering
    \includegraphics[width=0.9\linewidth]{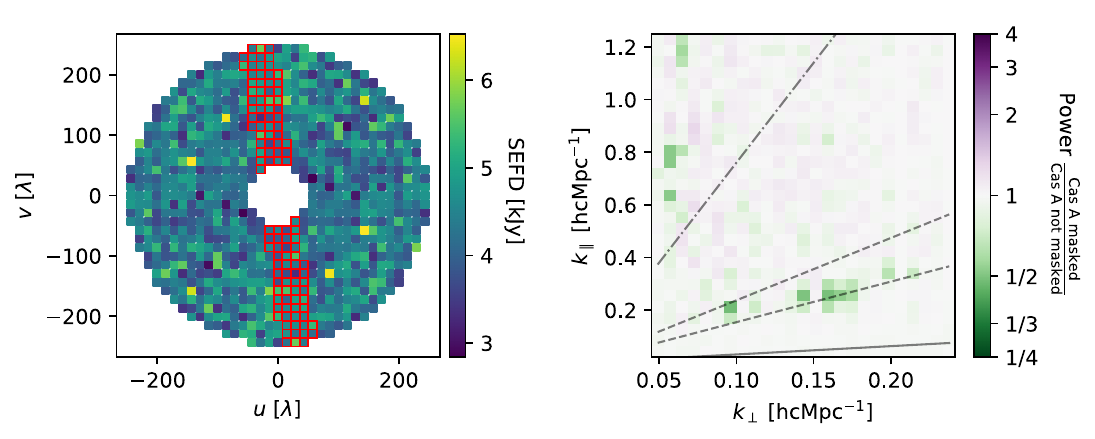}
    \caption{The effect of flagging Cas A in simulation \texttt{r5a}. Left: the SEFD computed using the spectrally differenced residuals after subtraction of all sources. Masked pixels are highlighted in red. Right: the ratio of a cylindrically averaged power spectrum with $u\varv$-space masking of {Cas A} and a similar power spectrum without masking {Cas A}. Both power spectra used in this ratio are a residual after GPR has been performed on the gridded data cube.}
    \label{fig:results_casflag}
\end{figure*}
To investigate how well GPR can capture the residuals of Cas A, we test the effects of $u\varv$-space masking. We use the same gridded visibilities but do not perform $(u,\varv)$-space flagging before applying GPR for one case while using the regular pipeline including $u\varv$-space masking for the other. The ratios between the results for simulation \texttt{r5a} are shown in \cref{fig:results_casflag}. We use this simulation rather than \texttt{r10a}, because the residuals of Cas A are stronger such that the effect of flagging is more visible.\par
On the left panel of \cref{fig:results_casflag}, there is no clear residual of Cas A visible, suggesting that the error is not dominating the error budget before flagging. On the right panel, flagging removes power in two regions: one near the expected delay of Cas A, and the other in a small region near $k_\perp = 0.06~h\mathrm{cMpc^{-1}}$ at high $k_\parallel$. We attribute the power reduction in the latter region to having very few baselines at these $k_\perp$-values. Flagging the $u\varv$-cells at which Cas A is expected to dominate may lead to a significant loss in the number of visibilities going into these $u\varv$-cells, leading to the loss of power here. We do not expect this to be an ionospheric effect, because the region is affected on a similar level for simulation \texttt{r10a}, which has weaker ionospheric errors than the shown simulation \texttt{r5a}.\par
In the region where Cas A is expected, the change in residual power upon flagging does depend on the ionospheric activity. We conclude that at least for some ionospheric activity levels, far-field source removal is not completely effective, even if both the sky and the beam are perfectly modelled. We expect such far-field sources to be more difficult to remove during GPR model-fitting as the frequency-dependent beam sidelobes and nulls in the far-field imprint a spectral structure on the source during the observation (see e.g. \citealt{cook_calibration_2021}). GPR mainly removes spectrally smooth power, to avoid subtracting the 21\nobreakdash-cm signal, such that these residuals are not removed. We expect the same to be true for other A-team sources, because they are also bright and far from the target direction. However, as shown by \citet{gan_statistical_2022}, Cas A and Cyg A are by far the most dominant far-field sources in LOFAR HBA observations of the NCP, and therefore dominate the off-axis foreground power \citep{cook_investigating_2022}. We analyse the spectral impact of the beam on bright far-field sources in more detail in future work.\par

\subsection{Amplitude and phase error transfer}\label{subsubsec:results_steps_partial}
\begin{figure*}
    \centering
    \includegraphics[width=\textwidth]{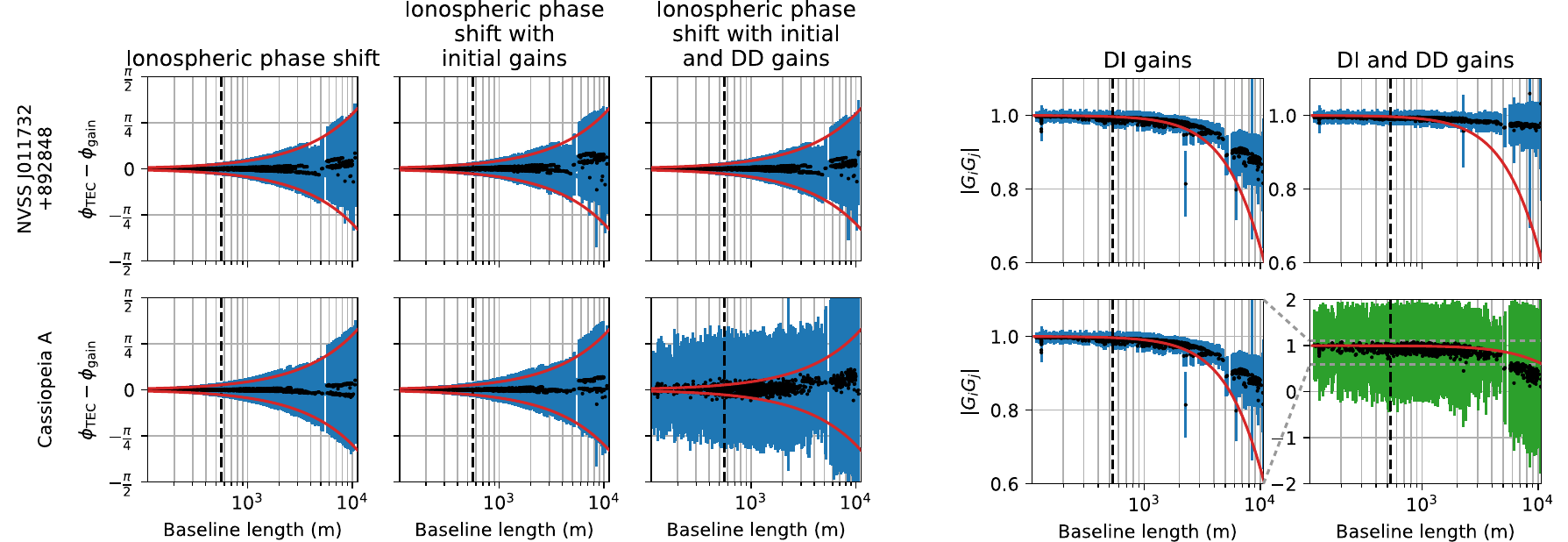}
    \caption{Difference between the ionospheric distortions and baseline gain at 140.3~MHz at different stages of data processing as a function of baseline length. Top row: bright calibrator source NVSS~J011732+892848 (also visible in the images of \cref{fig:results_images}), bottom row: Cas A. From left to right the first three columns show: the phase shifts of the TEC in the direction of the source (without including gains), the difference between the ionospheric phase shifts and the initial gain solutions, and the difference between the ionospheric phase shifts and the combination of both initial and DD-phase solutions. The right two columns show the amplitude levels of the gain solutions for initial calibration baseline gains, and the combination of initial and DD-gains respectively. The black dots represent the median value across time, and their blue error bars show the standard deviation. These points split apart for longer baselines because of sign changes due to baseline orientation. The bottom-right panel is plotted in a different colour and on a different scale. The vertical black dashed lines show the $250~\lambda$ baseline length. Finally, the red lines show the expected behaviour for a source in the zenith at a 10~km diffractive scale.}
    \label{fig:results_gainvariance}
\end{figure*}
One of the main motivations for the current work is to assess whether ionospheric errors may lead to gain errors on long baselines during calibration, resulting in an error transfer to the short baselines used for power spectrum estimation. Because the solution intervals (being 30~s for the initial calibration, 10~min for the Cas A DD-calibration and 2.5~min for the phase centre DD-calibration step) are longer than the simulation time resolution, a decorrelation in the visibilities occurs. This results in a change in both the amplitude and phase of the inferred gains. To investigate this further, we analyse the estimated gains from initial and DD-calibration on a baseline level. For stations $i$ and $j$ with station gains $g_i$ and $g_j$, we define the baseline gain $g_i{g_j^*}$, where $g^*$ indicates the complex conjugate of $g$. Each pair of stations is used only once, such that if baseline $i,j$ is included, baseline $j,i$ is omitted. The differences between the ionospheric effects and gains for all baselines at 140.3~MHz (the central frequency in our simulation band) are visualised in \cref{fig:results_gainvariance} for a single polarisation. Two cases are shown here: off-axis source Cas A, and {NVSS~J011732+892848}, a bright source at 30~arcmin from the target direction (see \cref{fig:results_images}) that is also used as an amplitude calibrator for observed LOFAR EoR data. Both sources dominate the flux density in their respective DD-calibration source clusters.\par

\subsubsection{Behaviour of gain phases}
The left panels of \cref{fig:results_gainvariance} show the difference between the phase shift as a result of ionospheric dispersion and the phase correction performed during calibration. The values shown are statistics over time\footnote{Within the simulation, this means that the averages are taken with to a changing diffractive scale for Cas A, since its elevation changes over time.}. In the case of a perfect correction for the ionospheric phase errors, this difference would vanish. The split in the mean values occurs due to baseline orientation, a sign change occurs between a baseline oriented parallel to the movement of the frozen TEC screen compared to a similar baseline oriented anti-parallel to this movement. The red envelope describes the expected standard deviation of phase errors for a source in the zenith at $r_\mathrm{diff}=10$~km, namely $\sqrt{D_\phi(\mathbf{b})}$~rad.\par
The top row demonstrates that overfitting is not a significant issue for target field sources, because the distribution of the phase errors does not evolve significantly during subsequent processing steps. Although the mean values of the phase differences shift, the standard deviations remain at approximately the same level as given by the structure function, despite it being impossible to solve for the rapid temporal variations of the ionosphere on the shortest temporal scales. This implies that neither a large bias nor an increase in gain errors are imprinted during data processing. The observed errors are the direct result of the ionosphere itself on time scales that are too short to solve for on short baselines. For {Cas A}, on the other hand, the variance of the gain solutions drastically increases in the DD-calibration step on all baselines. \par
We conclude that this occurs when Cas A passes through nulls in the station beams. When this happens, there is very little signal power to constrain the gain. Due to this low signal-to-noise ratio, some stations obtain gains that deviate strongly from the true values. In these short intervals the overall standard deviation increases. The visibilities at these nulls do not dominate the residual power of Cas A, because in or near the null, the erroneous gains are strongly attenuated by the synthesised beam as the source traverses a null during visibility prediction. The overall impact of the off-axis sources on the power spectrum is therefore limited when they traverse a beam null \citep{cook_investigating_2022}. \par

\subsubsection{Behaviour of gain amplitudes}
The right-hand panels in \cref{fig:results_gainvariance} show the gain amplitude per baseline averaged over time. Because the ionosphere does not directly influence the amplitude of incident radiation in our simulations, these amplitudes approach unity in the ideal case. However, the ionosphere does indirectly influence the observed source amplitudes. This is due to uncorrected phase errors due to the ionosphere, which leads to `seeing' determined by the phase structure function around these sources and a lowered peak flux compared to the original point source. The flux is spread out more and the speckle noise increases random errors in the data. 
The seeing-halo is given by $\left< V(\mathbf{b}) \right>=V(\mathbf{b})\mathrm{exp}\left[-\frac{1}{2}D_\phi(\mathbf{b})\right]$ \citep{koopmans_ionospheric_2010,2015MNRAS.453..925V}, where $\left<\cdot\right>$ indicates the expectation value when averaged over sufficient time or well-separated baselines, and $\mathbf{b}$ a baseline vector. The shape of this amplitude drop-off is shown in red on the right side of \cref{fig:results_gainvariance}.\par
To compensate for the lower peak flux compared to the sky model as a result of this `seeing', the amplitude of the average gain is lowered during calibration. This effect is stronger for longer baselines, because they experience larger ionospheric errors. In the direction of {Cas A}, the variance of the amplitudes is very large, due to the aforementioned issues near nulls.\par

\subsubsection{Gain phases and amplitudes in the power spectrum}
\begin{figure}
    \centering
    \includegraphics[width=\linewidth]{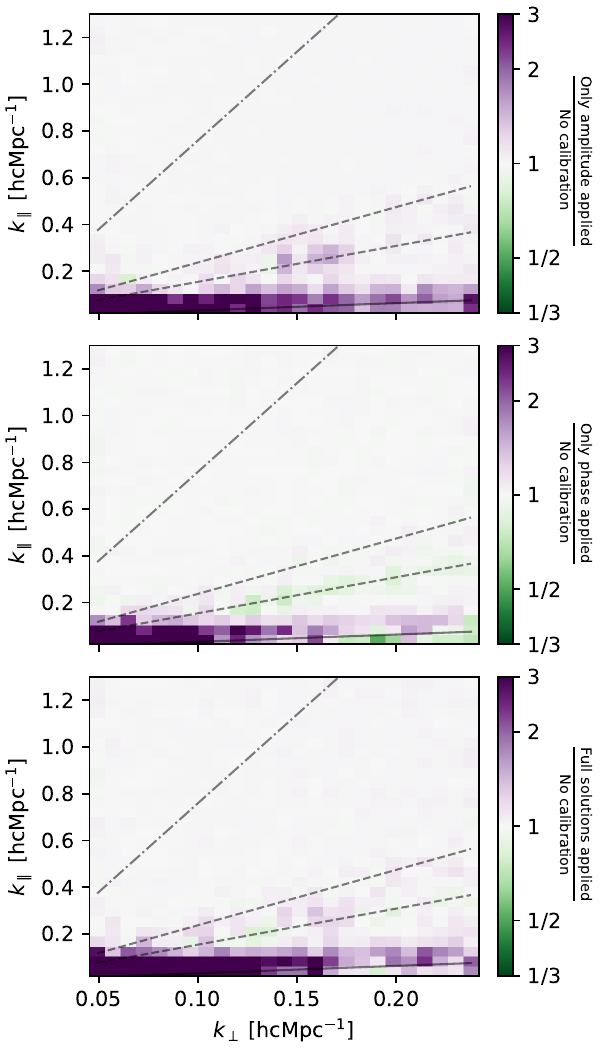}
    \caption{The ratio between the correction and sky-model subtraction using only part of the gain solutions and an uncalibrated residual computed using gains of unity for sky-model subtraction. We perform both initial calibration and DD-calibrated source subtraction. Top: only amplitude solutions are used, middle: only phase solutions are used, bottom: both phase and amplitude solutions are used (as in the regular pipeline). Amplitude solutions without phase solutions show an increased residual throughout the wedge. Phase solutions without amplitude show a slight reduction in power near the delay of Cas A and at the largest $k_\perp$ at the lowest $k_\parallel$, but an increase at the lower $k$-modes. A combination of both parts of the gains behaves like a sum of both excesses.}
    \label{fig:results_partial}
\end{figure}
To test the effects of both types of corrections separately, we compute cylindrically averaged power spectra using only phase and only amplitude corrections respectively. We do not recompute the gains, but instead apply either the phase or the amplitude part of the solutions. We do this both for the gain correction during initial calibration and for source subtraction during DD-calibration. \cref{fig:results_partial} shows the results for simulation \texttt{r10a}. The top panel shows the ratio between a power spectrum created using only amplitude corrections and a power spectrum created using gains of unity. The middle panel shows the same, but for phase instead of amplitude and the bottom panel shows the full gains versus gains of unity.\par
In the top panel, the ratio only deviates from unity inside the wedge, where foregrounds and their mode-mixing effects reside. All deviations are larger than unity, indicating that, in the absence of instrumental errors, amplitude corrections are detrimental to foreground removal. In the modes strongly affected by A-team flagging (the wedge area of Cas A around $k_\perp=0.15$ -- $0.19~h\mathrm{cMpc^{-1}}$), we see increased errors due to amplitude solutions, but not due to phase solutions. In the plot which uses only phase solutions (middle panel), stronger residuals are introduced at low $k$-modes. However, a slight reduction in residual power appears in other places in the wedge. This coincides with the larger $k_\perp$ modes on the delay lines of the NCP source and {Cas A}, indicating that in some calibration directions, the phase corrections reduce the residual power. Overall, however, we still see an excess as a result of phase calibration. We conclude that we are only able to solve for the ionosphere in the phase part of the gain of some of the longest baselines in power estimation, but are unable to do so for most of the power spectrum. Furthermore, when gain amplitudes are also computed, we cannot solve for the errors on any power-spectrum baselines anymore.\par

\subsection{Time coherence}\label{subsec:results_timecorr} 
\begin{figure*}
    \centering
    \includegraphics[width=\textwidth]{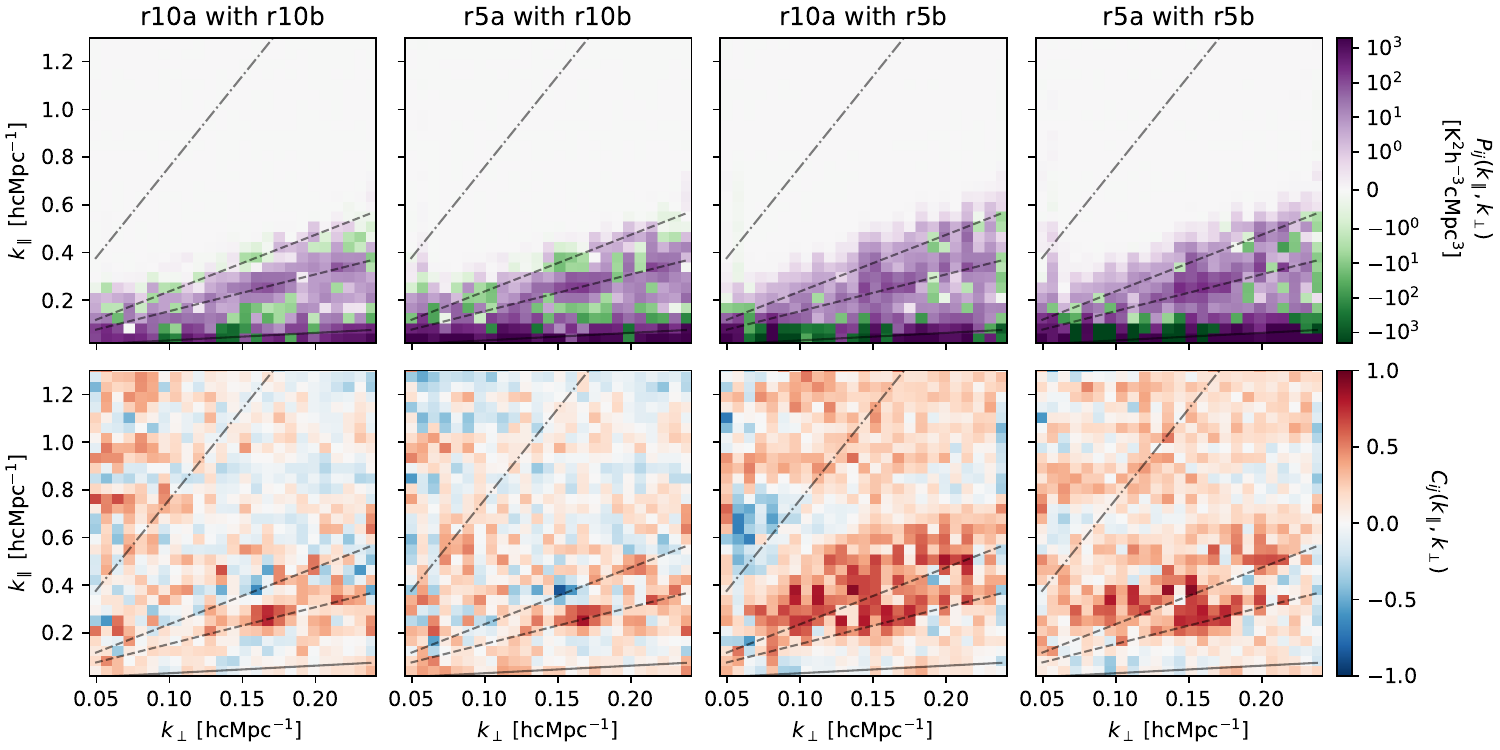}
    \caption{Cylindrically averaged cross-power spectra (top) and cross-coherences (bottom). Left: Correlation between simulations \texttt{r10a} and \texttt{r10b} (both with $r_\mathrm{diff}=10$~km, but at a different LST). Second column: correlation between simulations \texttt{r5a} and \texttt{r10b}. The TEC screens are the same as on the left, but the TEC-values of \texttt{r5a} have been scaled to have $r_\mathrm{diff}=5$~km. Third column: correlation between simulations \texttt{r10a} simulation and \texttt{r5b}. These simulations also have diffractive scales of $r_\mathrm{diff}=10$~km and $r_\mathrm{diff}=5$~km, like in the previous column, but these are at the same LST, such that the baselines observe the same sky. Final column: Correlation between simulations \texttt{r5a} and \texttt{r5b} (both with $r_\mathrm{diff}=5~$km and at the same LST). Cross-power in all plots is concentrated in the foreground wedge, approximately up to the upper delay line of {Cas A}. In the cross-coherence, the lower part of the wedge shows a weaker signal, indicating that the cross-coherence levels in this area are low, despite the cross-power being high. There is no clear difference between the two left-most columns, whereas the right-most columns show a larger coherence. This supports the idea that the coherence between different simulations is not dependent on the ionosphere itself, but more so on LST.}
    \label{fig:results_cross}
\end{figure*}
The time coherence of residuals after calibration is one of the most important metrics to determine whether the LOFAR EoR data processing is successful. A low or absent time correlation indicates that integrating more data will lead to a reduction in excess variance, and therefore make a detection of the 21\nobreakdash-cm-signal possible. A caveat to this is that an uncorrelated error on top of the thermal noise can still make the required time integration so large that it becomes prohibitive. If strong time correlations are present, however, any residual power from the foregrounds could pose a systematic limit on the detectable signal power. On the other hand, strong correlation may make it easier to model the residual foregrounds such that they can be removed more effectively. \par
To gauge the correlation between foreground residuals in different MSs, two metrics are used: the cross-power and the cross-coherence. The cross-power between simulations $i$ and $j$ is computed as
\begin{equation}
    P_{ij}(k_\perp, k_\parallel)=\mathbb{V}_c\left<\tilde{I}^*_i(u,\varv,k_\parallel)\ \tilde{I}_j(u,\varv,k_\parallel)\right>\ \mathrm{[K^2h^{-3}cMpc^3]}.
    \label{eq:results_crosspower}
\end{equation}
Here $\mathbb{V}_c$ is the observed comoving cosmological volume, $<\cdot>$ indicates a cylindrical average in $(u,\varv)$-space to transfer to cylindrically averaged coordinates, and $\tilde{I}(u,\varv,k_\parallel)$ refers to gridded brightness temperature cube at voxel $(u,\varv,k_\parallel)$. We also use the cross-coherence between simulations $i$ and $j$, 
\begin{equation}
    C_{ij}(k_\perp, k_\parallel)=\frac{\left<\tilde{I}_i^*(u,\varv,k_\parallel)\tilde{I}_j(u,\varv,k_\parallel)\right>}{\sqrt{\left<|\tilde{I}_i(u,\varv,k_\parallel)|^2\right>\left<|\tilde{I}_j(u,\varv,k_\parallel)|^2\right>}}.
    \label{eq:results_crosscoherence}
\end{equation}
Since the cross-coherence is normalised by the power spectra, its value is between $-1$ and 1 and its magnitude is large if there is a strong (anti-)correlation, regardless of the power level. We determine both the cross-power and cross-coherence to illustrate not only the level of correlation ($C_{ij}$) but also the impact that would have on a power-spectrum measurement ($P_{ij}$).\par
We have not included the residual foreground removal step in this analysis, because the residuals after GPR are already below the thermal noise floor. Additionally, we do not perform $u\varv$-space masking, to be able to interpret how the residual of Cas A behaves. Finally, thermal noise has not been included in this analysis. The reason for this is, first of all, that realistic thermal noise is uncorrelated over time, and will therefore artificially decrease the coherence and increase the sample variance. Furthermore, we have chosen to perform several simulations with the same thermal noise realisation to avoid random realisation differences between simulations. This means that the thermal noise would dominate the cross-power between these simulations if it is not removed. \par
The noiseless cross-spectra are shown in \cref{fig:results_cross}. In the first two columns, we show the correlation between two different LST ranges. The diffractive scales in the left-most column are 10~km for both simulations, whereas the observation in the middle column has a scaled TEC screen with $r_\mathrm{diff}=5~\mathrm{km}$ (but otherwise the exact same ionospheric realisation). In both simulations, the cross-coherence has a peak at the delay of {Cas A} around $k_\perp=0.15$ -- $0.19~h\mathrm{cMpc}^{-1}$. The location of this peak is comparable between the two different ionospheric activity levels and coincides with the pixels where A-team flagging has a large effect (see the right panel of \cref{fig:results_casflag}). The peak in the cross-power is slightly higher for the higher ionospheric activity level. For real data, this means that after combining nights, some correlation between different observations is expected to be present at the top of the wedge and the coherent power level is expected to depend on the residual power level in the individual observations. However, as A-team flagging and GPR are still performed after the steps we show here, this correlated power is likely removed later. \par
The third and fourth columns, which show the cross-coherence between two simulations at the same LST and date, show a large increase in cross-coherence at delay modes near and above Cas A. The third column has the same diffractive scales as used in the second column, whereas the fourth column uses the simulation with the same ionospheric realisation at a higher activity level again. Interestingly, the correlation region is narrower in the fourth column, even though both simulations in this spectrum use the same diffractive scale. The opposite would be expected from ionospheric speckle noise, because the `seeing-halo' around the sources would take the same form at the same LST and diffractive scale, whereas a higher diffractive scale would lead to a narrower halo and therefore less correlation. Therefore, we attribute the increased correlation for the same LST to a gain bias introduced during calibration. This is most likely a gain amplitude-based effect, because the gain amplitudes cause an increase in the foreground residuals in the relevant parts of the power spectra (see \cref{fig:results_partial}).\par
We conclude that the coherence between different simulations is more dependent on LST separation than on ionospheric activity level. Simulations at different LST mostly have a high coherence at pixels that are strongly affected by A-team sources, that will be masked out later in the pipeline. As such, we expect ionospheric errors to be removed when combining a sufficient number of nights at different LSTs.

\section{Summary and Conclusions} \label{sec:disc}
In this work, we analyse the impact of ionospheric effects on the 21\nobreakdash-cm signal power spectrum extracted from simulated LOFAR HBA data sets that are processed in a very similar way to real observations. To do so, we create simulations that isolate ionospheric errors from other possible sources of error and process these simulated data sets using the LOFAR EoR data analysis pipeline. This data processing pipeline includes initial gain-calibration, multi-direction gain calibration for source subtraction, $u\varv$-space masking of bright A-team sources, and residual foreground removal through Gaussian process regression \citep{mertens_improved_2020}. Crucially, these steps are performed using the same calibration settings as used on real data, including a 50 -- 250~$\lambda$ and 250 -- 5000~$\lambda$ baseline split in which longer baselines are used for calibration and shorter baselines for 21\nobreakdash-cm signal estimation. Our simulations of a typical LOFAR HBA dataset of the North Celestial Pole at $z=9.1$ include ionospheric phase errors representing one nominal, and one more extreme case, at the limit where real data is typically discarded. We conclude the following: 
\begin{enumerate}
    \item\noindent\textit{Ionospheric errors that occur under normal observing conditions with LOFAR HBA ($r_\mathrm{diff}>5~$km) are not a dominant component of the excess power as seen in the 21\nobreakdash-cm power spectrum.} Unlike the results for a single night of real observations shown by \citet{mertens_improved_2020}, we are able to remove the foregrounds to below the thermal noise, despite having a lower diffractive scale \citep{gan_assessing_2023}\footnote{The pipeline used in this work is slightly different, however.}. After GPR, no excess variance remains at a level exceeding ten per cent in any of our simulations, considerably lower than the excess variance of factors a few in current observations. We conclude that the propagation of ionospheric errors from longer baselines, which are the most affected by induced phase errors, to shorter baselines, is not a dominant contributor to the excess variance.\\
    \item\noindent\textit{The ionosphere makes signal extraction more difficult.} We have shown that configurations that remove the ionospheric errors exist. However, our configuration shows a positive noise bias of a few per cent, and the fitted component for mode-mixing varies on spectral scales that could approach those of the 21\nobreakdash-cm signal itself. This can potentially cause signal suppression, so more extensive tests of the GPR settings, or more physically motivated foreground kernels, are needed in the case of real data. \\
    \item\noindent\textit{The gain solutions that try to correct for ionospheric errors imprint low-level coherent errors onto the data after sky-model subtraction.} This even occurs when datasets with uncorrelated ionospheric phase screens are used. This is mostly likely due to the bias in the gain solution as a result of the decorrelation of the visibilities within a solution interval.\\
    \item\noindent\textit{Removal of bright off-axis sources remains more difficult than sources in the target field (the North Celestial Pole).} 
    Errors remain present near the delay of Cas A in our simulations. For a smaller (5~km) diffractive scale, $(u,\varv)$-space masking is needed to remove the source down to the thermal noise level. Furthermore, the gains computed in the direction of this source show a high variance, possibly due to the crossing of nulls and side lobes of the primary beam. We show that these residuals also occur when the modelling and calibration beams are identical and therefore devoid of beam modelling errors. Although these subtraction errors are exacerbated by the ionosphere, we cannot completely attribute them to ionospheric errors, as they also occur in our simulation without an ionosphere. 
\end{enumerate}
Based on our findings, there is no need to exclude ionospheric activity levels currently used in processing. However, a physically motivated ionospheric covariance kernel may need to be added to our Gaussian process regression step as higher sensitivity levels are approached, both for LOFAR and the future SKA. However, at the moment, the imprint of the ionosphere is well below that of the thermal noise after full data processing including Gaussian process regression. Therefore, the ionosphere is not the cause of the observed excess variance in LOFAR HBA 21\nobreakdash-cm signal datasets. It might be possible to improve the multi-direction calibration step by taking speckle noise into account during forward prediction, such that gain amplitude suppression is avoided, especially in the case of off-axis sources such as Cas A. Because the sky contains several bright off-axis sources, the higher sensitivity of SKA may lead to many off-axis sources being detected above the noise level. It then becomes infeasible to mask all such sources, because this can lead to a significant loss of $u\varv$-coverage. These improvements may be especially useful for future processing with SKA, which has much higher sensitivity levels and may therefore suffer from ionospheric errors to a larger extent, although this may be mitigated somewhat by the better $u\varv$-coverage of SKA. \par

\section*{Acknowledgements}
SAB, LVEK, JKC, BKG, SG, CH and SM acknowledge the financial support from the European Research Council (ERC) under the European Union’s Horizon 2020 research and innovation programme (Grant agreement No. 884760, “CoDEX”). EC acknowledges support from the Centre for Data Science and Systems Complexity (DSSC), Faculty of Science and Engineering at the University of Groningen. FGM acknowledges the support of the PSL Fellowship. 

\section*{Data Availability}
The simulation and data processing pipelines underlying this article are available at \url{https://doi.org/10.5281/zenodo.12188371}. The simulations underlying this article will be shared on reasonable request to the corresponding author.



\bibliographystyle{mnras}

\appendix
\section{Sky model}\label{app:sky model}
An excerpt from the supplementary sky-model file is tabulated in \cref{tab:app_skymodel}.  Note that the right ascensions of sources within a calibration cluster can have a large spread, as this does not necessarily lead to a large angular separation near the NCP. 

\begin{table*}
    \centering
    \caption{Excerpt from the file that is provided as supplementary material to this work. All listed numbers are given with the same sign and number of significant digits as used in the file itself. Because all sources in the sky model are unpolarised, flat-spectrum point sources, we do not show the superfluous fields for polarised components, spectral shape and characteristics of Gaussian sources.}
    \label{tab:app_skymodel}
    \begin{tabular}{lcccccccccccccc}
\hline
Type&calibration cluster&source label&RA&Dec&flux density (Jy)\\
\hline
 cluster definition& 3C61.1&& $+02^\mathrm{h}33^\mathrm{m}06\rlap{.}^\mathrm{s}8705$& $+86^\circ30'45\rlap{.}''2960$\\
 cluster definition& Patch\_0&& $+01^\mathrm{h}26^\mathrm{m}43\rlap{.}^\mathrm{s}8591$& $+89^\circ27'48\rlap{.}''4997$\\
 cluster definition& Patch\_1&& $-02^\mathrm{h}02^\mathrm{m}25\rlap{.}^\mathrm{s}5563$& $+89^\circ14'43\rlap{.}''1056$\\
$\hdots$&$\hdots$&&$\hdots$&$\hdots$\\
point source& 3C61.1&s0c125 &$+02^\mathrm{h}22^\mathrm{m}34\rlap{.}^\mathrm{s}186$& $+86^\circ17'19\rlap{.}''596$& 37.493542737027255\\
point source& 3C61.1&s0c199&  $+02^\mathrm{h}46^\mathrm{m}56\rlap{.}^\mathrm{s}127$& $+86^\circ18'30\rlap{.}''985$& 0.2866027935546541\\
point source& 3C61.1&s0c204&$+02^\mathrm{h}17^\mathrm{m}34\rlap{.}^\mathrm{s}851$& $+86^\circ44'34\rlap{.}''757$& 0.19606612284926137\\
point source& Patch\_0&s0c1063& $-02^\mathrm{h}35^\mathrm{m}35\rlap{.}^\mathrm{s}333$& $+89^\circ41'09\rlap{.}''287$& 0.03310936307446051\\
point source& Patch\_0&s0c1064&  $-02^\mathrm{h}36^\mathrm{m}52\rlap{.}^\mathrm{s}226$& $+89^\circ41'17\rlap{.}''052$& 0.27724967346183066\\
$\hdots$&$\hdots$&$\hdots$&$\hdots$&$\hdots$&$\hdots$\\
    \end{tabular}
\end{table*}

\section{Reference data-processing pipeline} \label{app:ref_pipeline}
The data-processing (or calibration) pipeline used in this work stems from the pipeline used to produce LOFAR EoR upper limits on the 21\nobreakdash-cm signal, as discussed by \citet{mertens_improved_2020} (see their Fig. 1 in particular). We have chosen to base our work on a recently updated version of this pipeline, which is used to produce an improved version of these upper limits by Mertens et al. (in preparation). The advantage of using this improved pipeline is that our results reflect the ionospheric impact on the most competitive LOFAR EoR results. However, as the improved pipeline is not yet described in a publication, we list the updates with respect to the pipeline presented by \citet{mertens_improved_2020} that are relevant to our work below:
\begin{itemize}
    \item The time interval in initial calibration has been increased to 30~s rather than 10~s.
    \item A-team flagging has been introduced after gridding. We describe how this is done in more detail in \cref{subsubsec:methods_cal_casflag}.
    \item The Gaussian process models used for residual foregrounds removal have been updated following the introduction of ML-GPR \citep{acharya_21-cm_2024,mertens_retrieving_2024}. As discussed in \cref{subsubsec:methods_cal_PS}, the kernels used in the analysis of our simulations cannot be the same as those used on real data. We therefore use a set of kernels that resembles those investigated in the improvement of the data processing pipeline.
\end{itemize}

\section{Error Evolution per simulation} \label{app:overviews}
\begin{figure*}
    \centering
    \includegraphics[width=\textwidth]{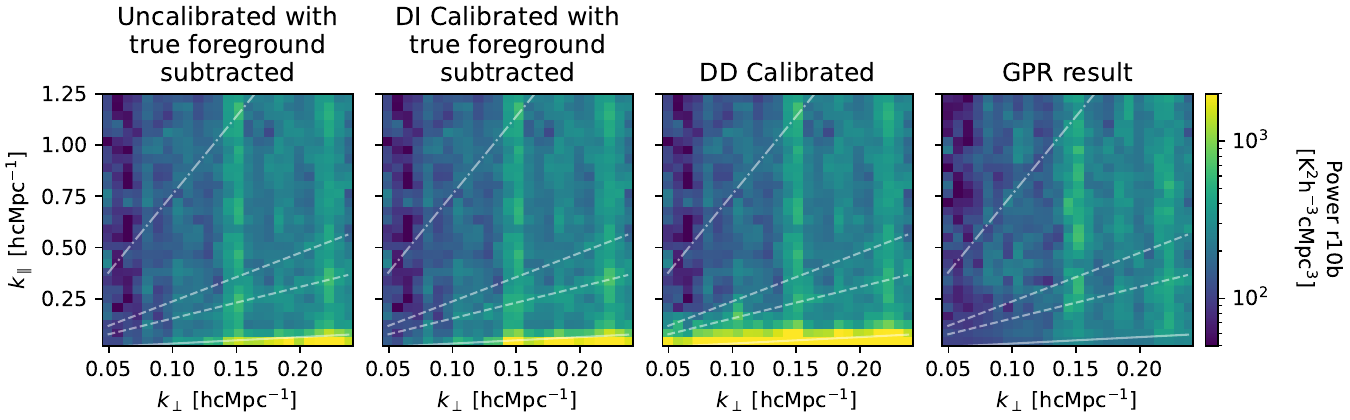}
    \caption{Overview of the residual power in cylindrical power spectra of simulation \texttt{r10b}. The columns represent the different calibration steps in the same way as \cref{fig:results_overzichtsplot10km}. The \texttt{TRUE VISIBILITIES} have been subtracted from the data in the first two columns, such that all columns represent residuals, even before the sky-model subtraction step in the pipeline.}
    \label{fig:app_overview_10km2}
\end{figure*}

\begin{figure*}
    \centering
    \includegraphics[width=\textwidth]{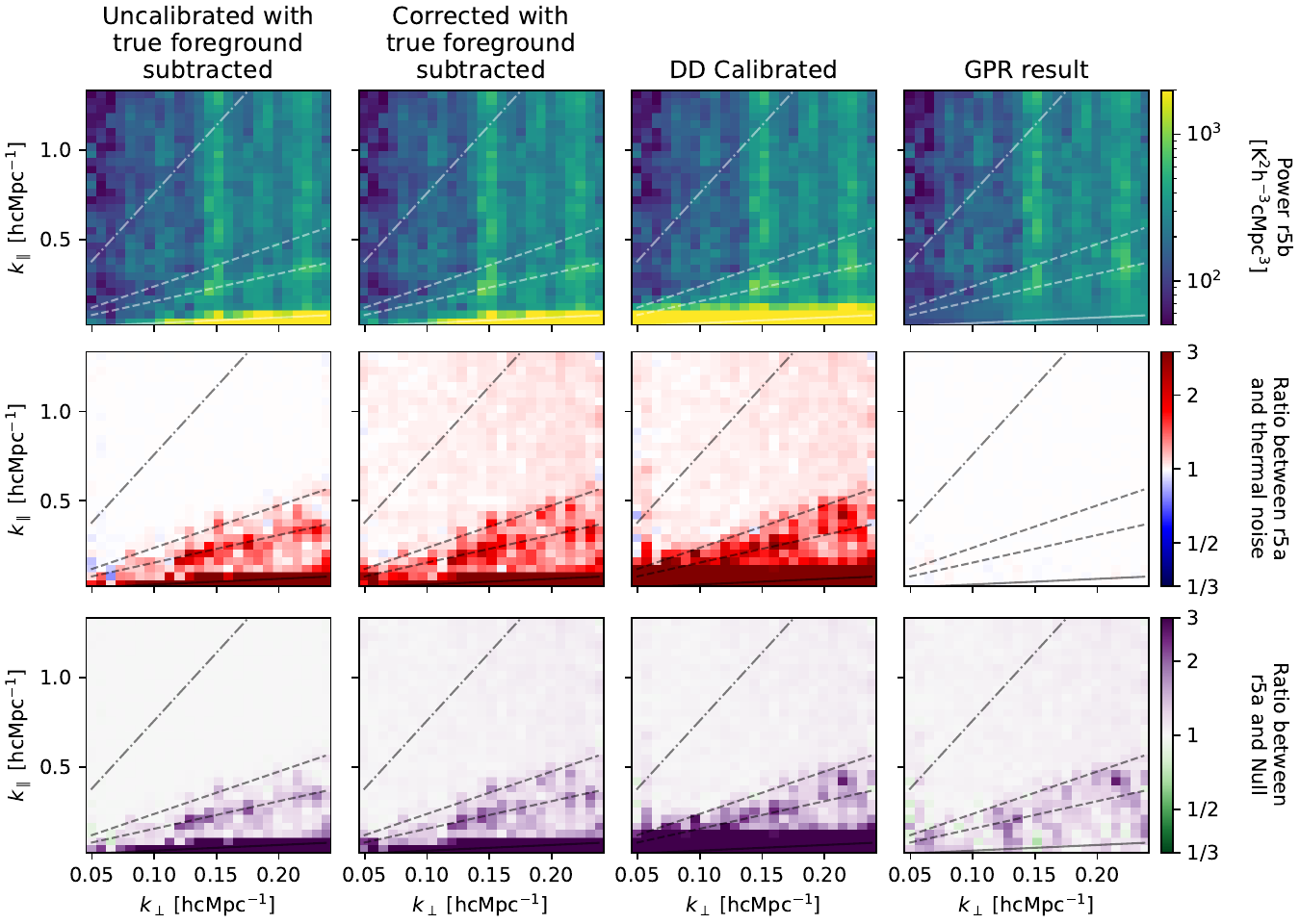}
    \caption{The same as \cref{fig:results_overzichtsplot10km}, but for simulation \texttt{r5a}. To make the comparison between these figures easier, we use the same colour scale. From top to bottom, the rows show the power spectrum of the residuals, the ratio between this power spectrum and the input thermal noise power spectrum and the ratio between the power spectrum and the same spectrum produced using the \texttt{Null} simulation. In the first three columns the input noise realisation is used as thermal noise, whereas in the last column, this is replaced by a thermal noise cube drawn from GPR.}
    \label{fig:app_overview_5km1}
\end{figure*}

\begin{figure*}
    \centering
    \includegraphics[width=\textwidth]{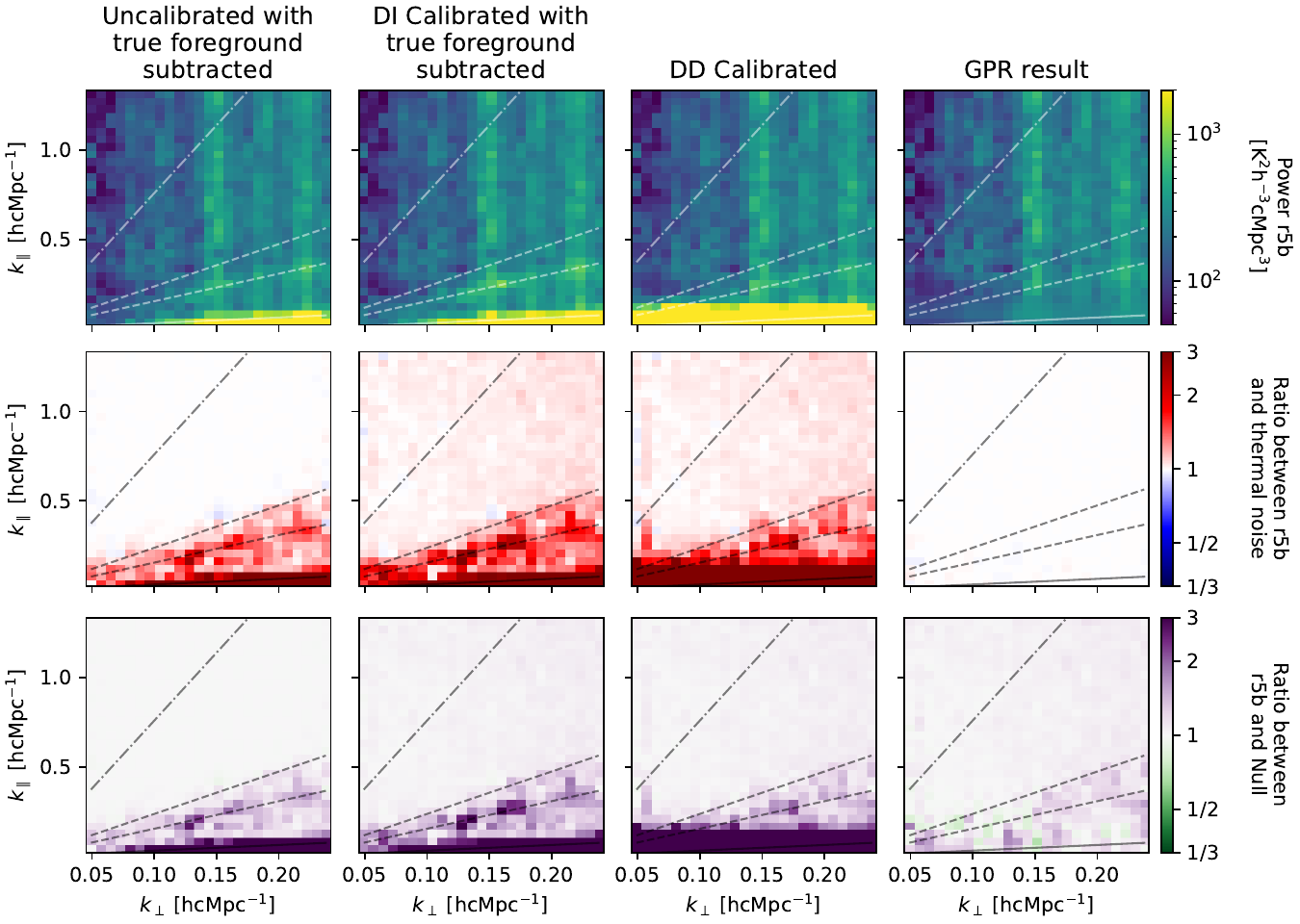}
    \caption{Overview of the calibration of simulation \texttt{r5b} simulation. The figure is plotted in the same way as \cref{fig:app_overview_5km1}.}
    \label{fig:app_overview_5km2}
\end{figure*}

Here, we provide propagation of ionospheric errors through the pipeline of simulations \texttt{r10b} (\cref{fig:app_overview_10km2}), \texttt{r5a} (\cref{fig:app_overview_5km1}) and \texttt{r5b} (\cref{fig:app_overview_5km2}). We cannot plot the ratio with the \texttt{Null} simulation and thermal noise realisation without performing additional simulations for \texttt{r10b}, because this simulation uses a different LST and a thermal noise realisation. As such, we only show the auto-power spectrum for that simulation. \cref{fig:app_overview_5km1,fig:app_overview_5km2} show a larger residual of Cas A in the ratio between the simulation with an ionosphere and the \texttt{Null} simulation. Similar to what is discussed in \cref{subsec:results_5vs10}, this effect is expected to be below the thermal noise level, and only visible here because the power spectra share the same LST and thermal noise realisation.

\section{Posterior distributions of residual foreground removal}\label{app:corner}
In this appendix, we show the posterior distributions from GPR run on the simulations with a 5~km diffractive scale. These simulations have a more active ionosphere than those shown in \cref{fig:results_corner10}, such that the foreground variances are higher (less negative on the log-scale).

\begin{figure}
    \centering
    \includegraphics[width=\linewidth]{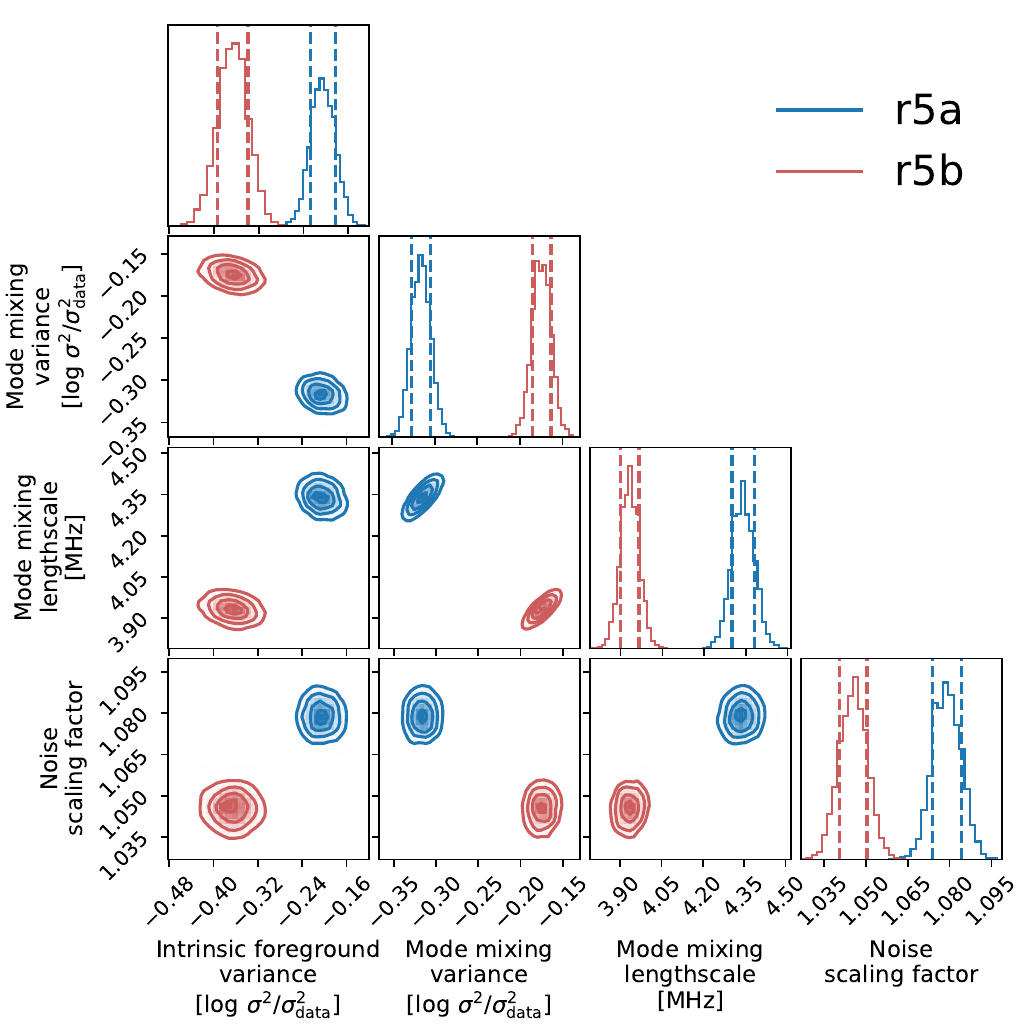}
    \caption{Corner plot of the posterior probability distributions of GPR on the $r_{\mathrm{diff}}=5$ km simulations. The vertical dashed lines in the marginalised distributions indicate the 68 per cent confidence level. }
    \label{fig:app_corner5}
\end{figure}



\bsp	
\label{lastpage}
\end{document}